# In the Mood to Exclude: Revitalizing Trespass to Chattels in the Era of GenAI Scraping

David Atkinson[1]

## Abstract

GenAI companies are strip-mining the web. Their scraping bots harvest content at an unprecedented scale, circumventing technical barriers to fuel billion-dollar models while creators receive nothing. Courts have enabled this exploitation by misunderstanding what property rights protect online. The prevailing view treats websites as mere repositories of *intellectual property* and dismisses trespass claims absent server damage. That framework grants AI companies presumptive access while ignoring the economic devastation they inflict. But the content is severable from the website itself. This paper reframes the debate: websites are *personal property* as integrated digital assets subject to the same exclusionary rights as physical chattels. When scrapers bypass access controls and divert traffic that sustains a website's value, they commit actionable trespass. The law need not create new protections; it need only apply existing property principles to digital space.

Courts and litigants have struggled to police unwanted, large-scale scraping because copyright preemption often narrows available claims, leaving copyright and its fair use defense as the primary battleground. Trespass to chattels offers a superior path, grounded in the fundamental right to exclude unwanted intrusions. Reviving this tort would protect not only content creators but also the digital ecosystem. Such protection would discourage exploitative scraping, preserve incentives for content creation, help protect privacy and personal data, and safeguard autonomy and expression. Reaffirming website owners' right to exclude is essential to maintaining a fair and sustainable online environment.

[1] Assistant Professor of Instruction, Business, Government and Society Department, McCombs School of Business, University of Texas at Austin, and Fritz Family Fellowship postdoctoral fellow at Georgetown University.

# I. Introduction

Your data is feeding artificial intelligence systems with insatiable appetites. This includes photographs you shared with friends, articles you labored to write, videos you produced, and any other data not secured behind a password-protected login (and some that is), even if it was intended for a certain audience at a particular time and in a specific context. Your words are tokenized, your images are processed, and your creative expressions are reduced to training or input data for AI models.

Web scraping bots collect this data. While the word "bot" may evoke an image of a physical robot, the bots at issue are software programs operating invisibly. They visit webpages, copy most of the content on each page, and repeat the process hundreds, thousands, or even millions of times. The goal is to collect as much data as possible for parsing, extraction, and incorporation into training datasets and prompts.[2]

Almost all of this scraping occurs without the permission or knowledge of those being scraped. Instead, these scrapers treat accessible information, including personal data and copyrighted works, as presumptuously fair game. The datasets developed through this process are used to train generative artificial intelligence (GenAI) models, including the chatbots, image generators, music generators, code generators, and video generators offered by OpenAI, Google, Meta, Microsoft, Amazon, Anthropic, and countless smaller organizations. The result is a profound asymmetry: AI firms accumulate competitive advantages worth billions, while creators face uncompensated exploitation of their work.

Legal resistance to predatory scraping has been swift but largely ineffective. As of January 2026, over 80 lawsuits challenge GenAI companies, with nearly all stemming from unauthorized content scraping.[3] Yet these efforts are consistently stymied by the doctrine of copyright preemption. By narrowing the legal battlefield to copyright and the associated 'fair use' defense, courts have inadvertently granted AI companies a presumptive right of access. This limitation serves GenAI developers at the expense of creators; if a developer prevails on a fair use argument, virtually any volume of unauthorized scraping becomes legally permissible. To break this impasse, we must look beyond intellectual property to the foundational principles of personal property.

---

[2] While most scraping on the web is for training data, about a fifth of the bots deployed by GenAI companies are to fetch information at the time the user asks a question. This is known as retrieval augmented generation (RAG). *See, e.g.,* Joao Tome, The Crawl-to-Click Gap: Cloudflare Data on AI Bots, Training, and Referrals, Cloudflare Blog, https://blog.cloudflare.com/crawlers-click-ai-bots-training/. (Tome also notes that only 18% of AI crawling was for search, and 2% was for user actions.)

[3] David Atkinson, AI Litigation Tracker – AtkinsonAILaw.com, https://docs.google.com/spreadsheets/d/1BgoFqE0C6148M2evzZU9A-WJN0oys7Pm/edit?gid=1305851266#gid=1305851266

Some scholars have argued that content on websites should be treated as real or personal property, with a right to exclude.[4] But others have noted that this analogy quickly breaks down because the content is not rivalrous.[5] When a scraper copies content, it does not deprive the original owner of the ability to continue using it. Courts have embraced this reasoning, concluding that trespass to chattels requires proof that scraping burdens the website's technical functionality. The courts also reason that, unlike when someone rummages through your backpack without permission (you are at least temporarily dispossessed of your property even if you are not, in fact, dispossessed[6]), scraping is not considered inherently harmful, and it does not typically dispossess the website owner of the website's content.

This analysis, however, commits a category error. The flaw lies not in property theory but in its misapplication. Courts tend to focus exclusively on website *content* while ignoring the *website itself* as an integrated digital asset. But there is no principled reason for bots to have presumptive access to content before we begin the legal analysis of whether they should be permitted such access. This oversight, replicated by plaintiffs who have internalized judicial skepticism, stems from misconceptions about what websites are and how scraping functions within the Internet's technical architecture.

The solution to protecting content from widespread exploitation lies in shifting focus from content to the website itself. Rather than expanding intellectual property protections, established principles of personal property apply directly to digital assets. It is largely uncontroversial that a website constitutes personal property. It is also widely accepted that people have a right to exclude others from their personal property.[7] Furthermore, no competing interest or legal principle justifies treating digital property as less protectable than its tangible counterparts.[8]

The prevailing skepticism toward digital trespass often rests on the 'nonrivalry' of information. Critics argue that because a scraper does not deprive the owner of their content, no harm occurs. This reasoning, however, ignores the rivalry of the website itself. Ownership of a domain and its associated server resources is inherently exclusive; two entities cannot exercise primary control over the same digital estate. Furthermore, while the server's physical condition may remain intact, the economic value of the chattel is severely impaired when GenAI scrapers divert traffic and bypass the value-exchange mechanisms (such as advertising or subscriptions) that sustain the property. Under the Restatement (Second) of Torts, impairment of a chattel's value is no less actionable than physical damage.[9]

---

[4] *See, e.g.*, Lawrence Lessig, Code and Other Laws of Cyberspace (1999).

[5] *See, e.g.* Daniel J. Solove & Woodrow Hartzog, *The Great Scrape: The Clash Between Scraping and Privacy*, 113 Cal. L. Rev. 1521 (2025). ("Property analogies break down because personal data is often shared, yet it is non-rivalrous, meaning when one person has it, it doesn't stop others from having it (or keeping it) as well.")

[6] *E.g.,* if someone were to rummage through your backpack while you were wearing it, so that you never lost control of the backpack, it would still likely constitute trespass to chattels.

[7] Moreover, the right to exclude is about controlling access and must not be confused with a right to expel or obstruct someone from a website after they *already* have access.

[8] This paper is emphatically *not* about criminalizing scraping, or the Computer Fraud and Abuse Act more generally. The focus here is on the tort of trespass to chattels as a remedy for violating a property owner's right to exclude.

Moreover, access to websites and their content is excludable because websites are capable of effectively controlling who can access their content.[10] If it is unlawful to rummage through a backpack (personal property) to copy content like a written poem that may be inside (even though the copying renders the act nonrivalrous because the owner retains the original poem) simply because the backpack's owner permits certain individuals (friends, spouses, children, etc.) to access it, the same concept should apply to websites and their content.[11]

Additionally, while courts have almost exclusively focused on whether the interference with chattel dispossessed the owner or impaired the chattel's condition or quality, they overlook that under the Restatement (Second) of Torts, affecting the *value* of the property can also support a trespass claim *even when the physical condition of the chattel is unaffected*.[12] The scraping of yesteryear may not have threatened the existence or economic viability of websites, and therefore their value, but the way GenAI companies scrape and the trend of people increasingly treating GenAI as an answer engine do.[13]

Re-empowering trespass claims would restore website owners' exclusionary rights that courts have eroded over the past two decades. Consequently, bypassing sufficient technological access controls designed to block bots (or anyone else) could give rise to legal claims that are currently unavailable under existing analytical frameworks.

Exclusion should not be treated as a game of cat and mouse, where the advantage lies with the scraper, who needs only to find one crack in the defenses, while the website owner must prevent all attacks from every angle. It should not matter if a bot can work around the most sophisticated blocking attempts. Doing so must still be unlawful, as is the case with all other forms of personal and real property. Just because someone can unzip a zipped backpack or enter the window of a house with a locked door does not make the trespass lawful.

---

[9] While Amazon.com is the domain name, it is also commonly and reasonably understood as the website of Amazon, the corporation. This paper will argue that the domain name should be treated as the property at issue.

[10] While being rivalrous and excludable may be necessary characteristics of property, they are not sufficient. In contrast, the right to exclude is a necessary and sufficient element of property.

[11] I will use a backpack for my analogies throughout this paper. Note that there is still debate over whether going through someone's bag without permission constitutes trespass. Some commentators insist there can be no trespass to chattels unless there is some harm, and mere rummaging may cause no harm. Still, the argument in this paper about websites is even stronger than the backpack analogy because there can be cognizable harm from GenAI scraping, as I'll explain in Sec. VI *infra*.

[12] *See* Restatement (Second) of Torts § 217 comments e and h (Am. Law Inst. 1965).

[13] A non-trivial amount of material on the web was posted with no expectation of pay, but also no expectation that it would be used as raw material for money-making ventures. Frequently, the expected compensation is notoriety and increased standing in various communities. Aside from the financial concerns, the co-opting of people's work also destroys the communities that created it in the first place.

This does not mean that scraping should be categorically illegal or that all sites should block all scrapers. Rather, website owners should have meaningful control over who accesses their websites. If an owner wants to block all bots, they should be able to do so, with legal recourse available. Similarly, if the website owner chooses to allow some or all bots to scrape their site, that should also be permissible.

Ultimately, reinforcing website owners' right to exclude would significantly benefit society. This paper does not support a project to make the Internet more closed or less accessible to most people or entities. In contrast to what some may fear, there is little reason to believe that empowering trespass claims would lead to an Internet that is more closed than the one we have today. In fact, I anticipate the opposite. Just as patents and copyright law facilitate the sharing of ideas, so too will enabling viable trespass claims. Most website owners who do not hide their content behind login accounts want to be accessible. But they also do not want to be exploited.

Such protection promotes the production and sharing of creative expression, including advances in the sciences and the arts, by ensuring creators can protect and control their creations. A robust exclusion right also provides an efficient and effective means of protecting and empowering First Amendment expression by combating the chilling effect that aggressive scraping can have on content creators who would rather not share speech than see it appropriated by GenAI companies. Finally, a strong right to exclude aligns with notions of equity, dignity, and autonomy by prohibiting the nonconsensual extraction and exploitation of creators' work.

This paper is organized into three parts. Part I lays the groundwork by examining the nature of property, the architecture of websites, and the mechanics of scraping in the GenAI era. Part II develops the legal argument for robust exclusionary rights, showing how trespass to chattels can be revitalized as a meaningful protection for digital property. Part III addresses limitations, exceptions, and counterarguments, and defends the proposed framework's practical implementation.

# Part 1: Property, Website, and Scraping Overview

## II. What is property?

Property is something you can own, though ownership rights vary significantly across legal systems and contexts. It was legal for humans to own other humans as property for centuries, for example. And while you cannot own the air around you, a nation can own its "airspace" and prevent other nations from entering parts of it.[14] This is another way of saying that property is a legal fiction, no more grounded in physical reality than contracts or torts. And, like contracts and torts, property law provides courts with plenty of flexibility to shape what it means to own property in ways that are most beneficial to society.

[14] *See, e.g.,* United States v. Causby, 328 U.S. 256 (1946)

## A. The Bundle of Rights

The typical conception is that property is something that comes with a "bundle of rights." This includes the rights of possession, use, alienation (i.e., the right to sell or give away), consumption, development, devising, transferring, protecting against state expropriation, pledging as collateral, and subdividing it into smaller interests. At bottom, property confers the authority to control a resource.

Among these sticks, the right to exclude is paramount. Scholars such as Thomas W. Merrill argue that the right to exclude is the *sine qua non* of property.[15] The United States Supreme Court has stated that the right of exclusion is "universally held to be a fundamental element of the property right,"[16] and "one of the most essential sticks in the bundle of rights that are commonly characterized as property."[17]

When considering the right to exclude, we must be careful not to conflate "exclusion" and "expulsion." Exclusion is the right to prevent others from accessing the property. Expulsion is the right to remove someone from the property *after* they have access.

The right to exclude will be the focus of this paper. As with all rights, it is not absolute. If a court believes the right conflicts with the Constitution or a statute, it may curtail the power of the owner.[18] Courts may also limit the right to exclude if it conflicts with the rights of others. This is where scraping enters the picture, and this paper will explore that supposed conflict.

## B. Policy-Based Reasons for Personal Property Law

There are three broad types of property: private, common, and public. Merrill succinctly describes them as follows:

> Private property may be said to exist where one person or a small number of persons (including corporations and not-for-profit organizations) have certain rights with respect to valuable resources. Common property may be said to exist where all qualified members of a particular group or community have equal rights

---

[15] Thomas W. Merrill, *Property and the Right to Exclude*, 77 Neb. L. Rev. 730 (1998). ("Give someone the right to exclude others from a valued resource, i.e., a resource that is scarce relative to the human demand for it, and you give them property. Deny someone the exclusion right and they do not have property… Whatever other sticks may exist in a property owner's bundle of rights in any given context, these other rights are purely contingent in terms of whether we speak of the bundle as property. The right to exclude is in this sense fundamental to the concept of property.")

[16] *Kaiser Aetna v. United States*, 444 U.S. 164, 179-80 (1979)

[17] *Dolan v. City of Tigard*, 512 U.S. 374, 384 (1994)

[18] *State v. Shack*, 58 N.J. 297 (1971)

> to valuable resources. An example would be a common pasture open to all members of a particular village for livestock grazing. Public property may be said to exist where governmental entities have certain rights with respect to valuable resources, analogous to the rights of private property owners. An example would be a municipal airport.[19]

Private property can include both real property, such as land, and personal property (i.e., chattels), such as jewelry. Contrary to what some scholars have claimed,[20] there are several policy reasons to support not only the idea of personal property in general, but also a personal property right with respect to websites specifically. For example, the right to own and benefit from property gives individuals a powerful incentive to work, save, and invest. When people know that the fruits of their labor will be protected, they are more likely to be productive, which in turn benefits the economy as a whole. This principle applies to websites, where creators invest substantial time, effort, and resources in developing and maintaining their digital properties. In contrast, if the works can be accessed and reproduced without consent or compensation, that disincentivizes the labor necessary to create such property.

Private property rights also establish clear ownership of resources, enabling their efficient use and exchange in the marketplace. This contrasts with common ownership, which is how scrapers often treat websites. Common ownership potentially leads to the "tragedy of the commons," where a shared resource is overused or neglected because no single person has a direct or sufficient stake in its long-term well-being.

A third reason to recognize a property right in a website is that, for some, a website represents a form of personal self-determination. The ability to own and control personal belongings, such as a website, provides a sphere of autonomy independent of both the state and other individuals. This enables individuals to make their own decisions about their lives and pursue their own objectives. Through the act of acquiring and using property, individuals can also express their will and establish their identity in the world. In this sense, owning something makes it an extension of oneself.

In short, the foundational rationales of property law apply directly to the digital realm. Websites are not an anomaly; they embody labor, value, and self-expression, and thus merit the same protections as other forms of personal property.

Having outlined the core principles of property law, the next step is to apply them to the digital realm by examining the nature and structure of websites as assets subject to ownership and exclusion.

---

[19] Thomas W. Merrill, *Property and the Right to Exclude*, 77 Neb. L. Rev. 730 (1998).

[20] Michael A. Carrier & Greg Lastowka, *Against Cyberproperty*, 22 Berkeley Tech. L.J. 1485 (2007). ("There is no tragedy of the commons, no need for incentives. There are no Lockean labor justifications. There are no Hegelian personhood rationalizations. Just as ominous, we conclude that the concept of cyberproperty is dangerous, unlimited, and unnecessary.")

# III. What is a Website?

Asking what a website is seems obvious and uninteresting. However, when pressed to define it concretely, people typically resort to some version of "I'll know it when I see it." Defining a website requires a specialized vocabulary that many people don't have and, understandably, don't need.

The specialized vocabulary is necessary here, though, because the legal argument and the correct application of the law require more than a superficial or vibes-based understanding of what a website is. For this paper, a website is a collection of interconnected web pages and related content (such as images, videos, audio, and documents) that share a common domain name and are published on at least one web server.[21]

In other words, a website is not just its content. Rather, it is the infrastructure that supports the components, along with the content. The content is only the visible and interactive elements of the website. Content is not the website itself, just as the words of a book are not the book itself.

## A. Server

The content of a website does not exist in some ethereal realm waiting to pop onto a screen when summoned by the correct keystrokes. Rather, it resides on a physical web server (often simply called a server), a specialized computer system that stores the website's files. When you visit a website, your browser sends a request to the server to provide access to the website and its content. The server is the "physical location" of the "library" where the website's "books" (files) are stored.

## B. IP Address and Domain Name

How do you connect to the server? Just as you may need an address to identify the location of a building or house, you also need an address to locate the server you're looking for. This comes in two forms: IP addresses and domain names.

The IP address (Internet Protocol address) is the numerical identifier used to route traffic to the server hosting the website (e.g., 1.1.1.1 for IPv4 or 2606:4700:4700::1111 for IPv6). Computers use IP addresses to identify and locate one another on the network.[22] An IP address functions like the precise GPS coordinates of the library. You could use the exact IP address to access a website by typing it in a search bar, but strings of numbers with no apparent relation to the website's content are challenging for humans to remember. This is where domain names enter.

Domain names are human-readable, memorable names that correspond to an IP address (e.g., google.com, example.org, mywebsite.net). When you type a domain name into your browser, the DNS (Domain Name System) automatically translates it into the corresponding IP address, allowing your browser to connect to the correct server. A domain name serves as the "street address" of the library.

---

[21] Wikipedia defines it as "any web page whose content is identified by a common domain name and is published on at least one web server." https://en.wikipedia.org/wiki/Website

[22] I'm mostly referring to the public internet, but there are other networks, like Local Area Networks (LANs).

A URL (Uniform Resource Locator) is the complete web address for a specific resource (such as a web page, image, or document) on a website. It includes the protocol (e.g., https://), the domain name, and often a path to the specific file or directory (e.g., https://www.example.com/about/team.html). The domain name is simply a component of the URL.

## C. Content

Once you connect to the website, you are on your way to accessing what you are really after: the content. Simply put, if you are seeing content, you are on the website.

One example of content is HTML (Hypertext Markup Language). This is the foundational language that structures a web page's content. It defines elements like headings, paragraphs, links, images, forms, and tables.[23]

Websites also use CSS (Cascading Style Sheets), a language that controls the presentation and visual styling of HTML elements. It dictates colors, fonts, layouts, spacing, and overall aesthetics.

Many sites also incorporate JavaScript. It's the programming language that adds interactivity and dynamic behavior to web pages. It enables features such as animations, form validation, interactive maps, server-side data fetching without page reloads, and more.

Finally, websites can include several types of media files, such as images (JPG, PNG, GIF, SVG), videos (MP4, WebM), audio (MP3, WAV), and other documents (PDFs, spreadsheets) that are embedded or linked on web pages.

## D. Reconceptualizing Websites

The breakdown above provides a clearer understanding of a website's boundaries and components. Perhaps the most important takeaway is that if a bot can extract content, it is already on the website. The legal question is whether unauthorized access—not copying—constitutes actionable property invasion.

# IV. How a Webpage Loads

Before conducting a legal analysis, it is helpful to first understand how webpages load. The process is far more complex than entering a URL and seeing the webpage appear on the screen fully formed. Rather, a series of extraordinarily fast steps occurs behind the scenes to display the website's content.

[23] If you right-click on a webpage and then click "Inspect" you can see the site's HTML.

Just as legal scholars refer to a GenAI supply chain,[24] we can usefully refer to a page-loading supply chain. Unfortunately, some technical jargon is unavoidable, necessitating numerous footnotes for those seeking a deeper understanding of the underlying processes. Fortunately, one does not need to understand every aspect of the process to comprehend the legal argument that follows in Part 2 of this paper.

I highlight steps 3, 4, 5, and 6 because they are key to this paper's argument.

1. Initiation of Navigation:
    - The process begins when a user types a URL into the browser, clicks a link, or reloads a page.
2. DNS Resolution (Domain Name System lookup):
    - The browser needs to determine the IP address of the server hosting the website.
    - It queries DNS servers[25] to translate the human-readable domain name (e.g., google.com) into an IP address (e.g., 172.217.160.142).
3. **Establishing a Connection (TCP/TLS handshakes):**
    - **Once the IP address is determined, the browser initiates a TCP three-way handshake[26] to establish a connection with the server. This ensures reliable data transfer.**
    - **If the connection is secure (HTTPS[27]), a TLS/SSL handshake[28] follows. This negotiates encryption parameters, verifies the server's identity, and establishes a secure channel for data exchange.**
4. **Sending the HTTP Request:**
    - **The browser sends an HTTP (or HTTPS) request to the server for the webpage's files (HTML, CSS, JavaScript, images, etc.).**
5. **Server Response:**
    - **The server receives the request, processes it, and returns the requested resources, starting with the HTML document.**
6. **HTML Parsing and DOM Construction:**

---

[24] Katherine Lee, A. Feder Cooper & James Grimmelmann, *Talkin' 'bout AI Generation: Copyright and the Generative-AI Supply Chain*, 72 J. Copyright Soc'y U.S.A. 251 (2025).

[25] These act as a translator for domain names into IP addresses. Think of them as the internet's phonebook.

[26] TCP stands for Transmission Control Protocol. The TCP three-way handshake is a three-step process used to establish a reliable connection between a client and a server on a TCP/IP network. It ensures that both devices are ready to send and receive data, and its primary purpose is to deliver data reliably, accurately, and in the correct order between a client and a server.

[27] HTTPS (Hypertext Transfer Protocol Secure) is a secure version of the HTTP protocol that uses encryption to protect communication between a web browser and a website. The "S" stands for "Secure" and indicates that all data exchanged is encrypted to prevent eavesdropping and tampering.

[28] This establishes a secure, encrypted connection between a client (like your web browser) and a server. It's the first step in using HTTPS and ensures that data is sent privately and securely. The term "SSL" is often used interchangeably with "TLS," but TLS (Transport Layer Security) is the more modern and secure successor to the older SSL (Secure Sockets Layer) protocol.

- **As the browser receives the HTML, it begins parsing it line by line.**
- **It constructs the Document Object Model (DOM), which is a tree-like representation of the HTML structure. Each HTML element[29] becomes a "node" in this tree.**

7. Fetching Resources and CSSOM Construction:
    - While parsing HTML, the browser identifies references to other resources, including CSS stylesheets, JavaScript files, and images.
    - It sends separate requests to download these resources.
    - As CSS files are downloaded, the browser parses them and builds the CSS Object Model (CSSOM), which represents the styles applied to the page. CSS is often "render-blocking," meaning the browser won't display anything until the CSSOM is complete.
8. Creating the Render Tree:
    - The browser combines the DOM and the CSSOM to build the Render Tree. This tree includes only the page's visible elements and their computed styles. Elements hidden by CSS are excluded.
9. Layout (Reflow):
    - Based on the Render Tree, the browser calculates the exact size and position of each element on the screen. This step determines the page's layout. If changes are made to the DOM or CSS later, this step may need to be repeated.
10. Painting (Rasterization):
    - The browser "paints" the pixels onto the screen based on the Render Tree and layout calculations. This is when the webpage's visual content becomes visible.
11. JavaScript Execution:
    - JavaScript files are downloaded and executed. JavaScript can modify both the DOM and the CSSOM, potentially triggering additional layout and paint operations.
12. Post-Load Interactions and Continuous Rendering:
    - After the initial page load, the browser continues to handle user interactions (such as clicks, scrolls, and form submissions) and dynamic content updates, which can trigger additional reflows and repaints.

As demonstrated above, a website does not simply exist in a completed state, awaiting a user to view it. Rather, accessing content requires several steps, each of which must be completed before proceeding to the next. Because the loading process is hierarchical, it allows for several potential points of intervention by website owners, third parties operating on the site owner's behalf, and third parties unrelated to the site owner.

This paper will refer to the steps outlined above throughout the legal analysis. Readers may find it helpful to note this page number for future reference, as it will aid in understanding how and when the legal concepts discussed below occur within the page-loading supply chain.

[29] A core component of an HTML document, used to structure and format content on a web page. Each element consists of a start tag, content, and an end tag, and it tells a web browser how to display a specific piece of content, such as a heading, a paragraph, an image, or a link.

With this technical foundation, we can now examine the practice of web scraping and its implications for property rights.

# V. Web Scraping

Now that we have covered what property is, what a website is, and how a website loads, there is one final explainer needed before we can fit all the pieces together: how scraping works. Web scraping is a software technique for automatically extracting data from websites. Scrapers deploy bots to extract content.

When scraping makes the news, it is often accompanied by claims of exploitation. Specifically, the content or website owner (which are not always the same entity) is upset because a bot has scraped the site without authorization and, in many cases, allegedly violated the site's terms of use. However, prior to 2022, scraping was a more niche issue, with only particularly egregious actors drawing significant attention. Many uses of scraping were initially considered beneficial, including creating datasets for academic research and enabling web search indexing, which makes search engines like Google possible. Other reasons to scrape include price monitoring, market research, lead generation, news aggregation, real estate analysis, and search engine optimization (SEO) monitoring.

## A. The Mechanics of Scraping

Scraping is a multi-step process that includes fetching, parsing, extracting, and storing information.

1. Scraping bots first must fetch the webpage's content. This typically involves making an HTTP GET request[30] to a URL (step 4 in the website loading supply chain).

2. The server's response (step 5) consists of the raw HTML (and potentially other assets if the scraper employs a more sophisticated headless browser[31]). At this point, the scraper gains possession of the website's content. When we colloquially refer to scraping, we typically mean copying content, and this is where it happens.

3. Next, the HTML is often parsed. This process converts raw HTML text into a structured, navigable format, such as a Document Object Model (DOM) tree. This makes it easier to locate elements of interest. The bot essentially "reads" the page's structure and content, enabling it to pinpoint the specific data it's programmed to extract (e.g., text from a particular div, image URLs, prices from a table).

---

[30] A method used to retrieve data from a web server. When you type a URL into your browser or click on a link, your browser sends a GET request to the server associated with that URL.

[31] A headless browser is a browser that is only useful for bots because it does not have visual elements. Because it functions like a typical browser it can fool sites into thinking a human is making the GET requests, not a bot. There are some valid uses for headless browsers, but they can also be abused by scrapers.

4. Then the scraper extracts the specific data it targets. Depending on the bot's purpose, this could include product names, prices, descriptions, reviews, contact information, dates, links to other pages, image URLs, and other relevant details.

5. Finally, the content is transformed from unstructured (as it appears in the webpage's code) into structured formats, such as JSON, CSV files, or SQL databases.

Modern AI-driven scraping can extract data from dynamic websites and multimedia content. Its ability to adapt to structural changes makes it substantially harder to block than traditional scraping methods. These advances also underscore the importance of allowing enforceable legal remedies before content is made available to visitors.

While the mechanics of scraping are often neutral, the manner in which it is conducted varies widely. This raises the question of what constitutes responsible versus problematic scraping behavior.

## B. Responsible Scraping Practices

Scrapers have developed an informal code of conduct to distinguish "responsible" scrapers from "problematic" ones. Compliant scrapers are less likely to have their IPs banned or face legal action. Best practices include:

1. Properly identifying the bot by using a descriptive name and following standard naming conventions for its user-agent.[32] For example, instead of using the generic user agent "python-scraper," use something more specific, such as "CompanyName-Bot."[33] This informs the site of the bot's origin and can imply its purpose.
2. Using Web Bot Auth integration or a declared list of IP ranges.
3. Providing contact information in case a website encounters an issue.
4. Serving a clear purpose, as described in sufficient detail on the scraper's public website.
5. Using different bots for various forms of web scraping, rather than imposing an all-or-nothing requirement on website owners.
6. Honoring a website's terms of service. These legal terms are often found in website footers. Some sites expressly limit or prohibit scraping.
7. Comply with the instructions in the robots.txt file. This is arguably the most established way for websites to communicate their scraping preferences. For example, the Wall Street Journal's robots.txt file is available at https://www.wsj.com/robots.txt. Reviewing the file shows that it blocks Perplexity's scraping bot.

---

[32] A user-agent is a string of text that a web scraper or bot sends as part of its HTTP request headers when it accesses a website. This string identifies the software making the request to the web server, much like a web browser's user-agent identifies it as a specific browser (e.g., Chrome, Firefox, or Safari).
[33] E.g., "Googlebot" from Google.

8. Use official APIs when available. APIs are designed for automated data access, are generally more stable, provide data in structured formats, and include clear usage policies and rate limits.
9. Prevent bots from overloading servers. One effective way to do this is to throttle requests to avoid patterns that resemble a denial-of-service attack.
10. If the scraper encounters an error, it should throttle further. A 429 error, formally known as "429 Too Many Requests," indicates that the scraper has exceeded acceptable request limits. This mechanism, called "rate limiting," is implemented by servers to prevent abuse, manage traffic, maintain fair resource allocation, and protect against potential attacks, such as Denial-of-Service (DoS) attacks.
11. When scraping a domain, scrapers should limit the number of concurrent requests to prevent server overload. This practice helps prevent server overload.
12. Scrapers can also run during off-peak hours. This usually happens at night in local time, when most people are asleep. Not only will the website have more bandwidth, but it is also less likely to affect other users' experiences.

It is worth emphasizing that the examples above are currently entirely voluntary.

## C. The Limits of Ethics

In practice, HTTP GET requests from bots are far from straightforward, contrary to what the responsible scraping guidelines listed above might suggest. Unscrupulous scrapers may actively circumvent blocking measures rather than respect website owners' clearly expressed preferences.

Instead, they employ tools like Browse AI and ParseHub that openly list "Extract data for LLMs" as a primary use case and feature advanced capabilities such as "human behavior emulation" and robust "bot detection, proxy management, automatic retries, and rate limiting" systems specifically designed to circumvent standard anti-scraping measures.[34] These tools may also employ user-agent spoofing and modify HTTP headers[35] to mimic legitimate user behavior.

The coexistence of sophisticated AI scraping tools designed to evade detection and content publishers' expanding adoption of blocking measures reveals an ongoing technological "arms race." AI companies and those developing tools for them continuously innovate to access data, while content owners develop increasingly advanced countermeasures.

As voluntary ethical norms prove insufficient to curb exploitative scraping, the rise of GenAI technologies has further intensified the challenges, necessitating a closer look at how GenAI companies engage in and benefit from large-scale scraping.

---

[34] *Browse.ai*, https://www.browse.ai/.
[35] HTTP headers are key-value pairs sent with HTTP requests and responses, providing additional context and metadata about communication between a client (e.g., a web browser) and a server.

# VI. GenAI's Nexus with Scraping

While GenAI bots function identically to search indexing bots at a technical level, the aggressive implementation of scraping bots by GenAI entities differs markedly from the scraping activities common on the Internet prior to 2022 in several ways, including their scale, frequency, and adherence to Internet norms that previously guided court decisions.

While it is sometimes argued that scraping increases exposure and may benefit content creators through greater reach, empirical data suggest otherwise in the context of GenAI. Referral traffic from GenAI applications is a fraction of that from traditional search engines, and news organizations have reported significant traffic declines since the rise of GenAI-powered search. These trends show that, rather than increasing exposure, large-scale scraping often erodes website value and undermines sustainable content creation. This Section will examine the impact of GenAI bots over just a few years.

## A. Why GenAI Companies Scrape So Much

While there are many legitimate reasons to scrape sites, in the age of GenAI, the most prolific scrapers primarily gather training data for GenAI models. Though web scraping has existed for decades, GenAI entities seek treatment comparable to that of historical scrapers, yet they represent a fundamentally different phenomenon. Their data consumption is more voracious and less discerning, and the previously mutually beneficial relationship between scrapers and society has all but disappeared.

GenAI companies require massive amounts of data to train their AI models. They ingest content, such as webpages, extract the desired parts, like text, and then convert that text into subwords, called tokens, to input into their models. This enables them to train the models to accurately predict which tokens are most likely to follow the preceding tokens. To illustrate the size of such datasets, Meta trained Llama 4 on "30 to 60 trillion tokens."[36]

The cavalier attitude GenAI scrapers adopt regarding scraping appears to stem in part from the permissive approach most sites traditionally maintained toward researchers and indexers. Put simply, because a website previously accepted being scraped for search indexes, there is an assumption that such websites should similarly accept scraping for GenAI training. Similarly, if academic scraping for research raised no legal concerns, the reasoning goes, then OpenAI, Meta, Google, and others should face few obstacles when pursuing comparable activities. This position finds support among scholars such as Edward Lee.[37]

---

[36] *Kadrey v. Meta Platforms, Inc., No. 3:23-cv-03417, 2025 WL 4123456 (N.D. Cal. May 1, 2025).* To help visualize, consider that one million seconds would be 11 days. One billion seconds would be about 32 years. One trillion seconds would be 31,688 years. Forty-five trillion seconds would be 1,426,000 years.

[37] Edward Lee, *Fair Use and the Origin of AI Training*, 63 Hous. L. Rev. 104 (2025), https://papers.ssrn.com/sol3/papers.cfm?abstract_id=5253011 ("This history should inform the courts' resolution of fair use in the copyright lawsuits against AI companies. Courts should evaluate whether the use of copyrighted works in AI training at universities serves a fair use purpose—or not—by examining

Historically, when non-consensual scraping occurred, such as competitive intelligence gathering, it typically targeted a relatively small portion of the web (i.e., large commercial websites that posed a competitive threat to companies with the resources and awareness to deploy scrapers and employ data scientists to analyze the scraped data). The same limitations applied to other scraping purposes, such as market research and search engine optimization analysis. Traditional scrapers mostly targeted strategically selected commercial sites rather than the entire Internet because universal scraping would yield a poor signal-to-noise ratio and be time-consuming and expensive.

GenAI entities pursue the opposite strategy by mostly indiscriminately collecting vast quantities of content. This represents a meaningful shift in not only the volume of data collected and the number of sites scraped, but also in the number of people and entities, large and small, that are negatively impacted by such ravenous scraping. In short, GenAI scrapers and historical scrapers are largely incomparable, and suggesting otherwise is misleading at best. Perhaps the most straightforward way to understand the impact is to examine recent data points.

## B. The Scale of Contemporary Scraping Activity

It's important to distinguish between the two primary types of GenAI scrapers[38]: those engaged in GenAI training and those conducting retrieval augmented generation (RAG) operations. Training-focused scraping involves bots that harvest virtually all public content on the Internet to form datasets used to train GenAI models. According to Cloudflare, these types of bots account for 80% of AI bot activity.[39] The RAG bots operate when a GenAI company deploys a bot to retrieve information in real time in response to user prompts. That data is then fed into the GenAI model as part of the prompt to enhance the model's output.[40]

---

whether AI training serves a different or transformative purpose. For, if the use of copyrighted materials by university researchers to develop AI models is copyright infringement and not fair use, then, a fortiori, the fair use defense of AI companies, commercial entities, must fail. Conversely, if the courts find that such university-based AI training has a legitimate fair use purpose, then courts should reject broad arguments that use of copyrighted works in AI training by companies cannot serve a fair use purpose—e.g., because it purportedly is "not transformative" at all.")

[38] While I often refer to GenAI companies doing the scraping, what I mean is that they are the cause of the scraping. They often scrape the Internet themselves, but they may also pay third parties to scrape content on the GenAI entity's behalf.

[39] Joao Tome, *The Crawl-to-Click Gap: Cloudflare Data on AI Bots, Training, and Referrals*, Cloudflare Blog, https://blog.cloudflare.com/crawlers-click-ai-bots-training/. (Tome also notes that only 18% of AI crawling was for search, and 2% was for user actions.) Other infrastructure provider figures vary. Fastly notes that "crawler traffic comprised 71% and fetcher, 29%." Fastly, Inc., *Fastly Threat Insights Report: From AI Crawlers to Headless Bots*, (Dec. 18, 2025), https://www.fastly.com/press/press-releases/fastly-threat-insights-report-reveals-organizations-are-fighting-back.

[40] According to TollBit, which seems to have a higher percentage of media companies as clients, "RAG bot scrapes now exceed Training bot scrapes across the TollBit network. From Q4 2024 to Q1 2025, RAG bot scrapes per site grew 49%, nearly 2.5X the rate of Training bot scrapes (which grew by 18%)1. This is a clear signal that AI tools require continuous access to content and data for RAG vs for training." *AI Scraping Is On The Rise: TollBit State of the Bots - Q1 2025*, TollBit (June 11, 2025), https://tollbit.com/bots/25q1/.

According to research by TollBit, a platform that facilitates content monetization by creating a marketplace where AI bots and data scrapers can obtain licensed access,[41] "Among sites with TollBit Analytics set up before January 2025, AI Bot traffic volume nearly doubled in Q1 [2025], rising by 87%."[42] Moreover, "Average scraping activity across the top 6 AI bots increased by 56% from Q4 to Q1…"

Cloudflare, an Internet infrastructure and security company significantly larger than TollBit, documented a similar surge in scraping, noting that scraping for GenAI training increased 65% in the first half of 2025.[43] This data helps explain Reddit's litigation claims that Anthropic scraped its site more than 100,000 times despite Anthropic's assurances that it would cease such activities,[44] why Anthropic scraped iFixit's website one million times over 24 hours,[45] and why even Wikimedia has raised concerns about scrapers overwhelming Wikipedia's servers with "65% of [its] most expensive traffic [coming] from bots."[46]

To provide context, it is helpful to compare AI bot activity with Google's established practices. Googlebot facilitates the indexing of the entire internet, enabling content delivery through Google Search. As the TollBit report notes:

> In Q2 2024, the scraping activity of the top 6 AI bots was roughly 10% the size of Googlebot's scraping activity. In Q1 2025, AI bot access to sites is now 60.29% that of Bingbot's activity, and 30.55% of Googlebot's total scraping.
>
> This data shows the dramatic increase in AI bot market share when compared to Googlebot and Bingbot activity. Google and Microsoft are companies that publishers have obviously had relationships with for 20+years. In one year [websites] are being hammered by new crawlers where the value exchange is no longer clear, especially if [the crawlers] don't drive traffic back to sites.[47]

Additional examples further illustrate the scope of the problem. Open source projects may be particularly vulnerable to AI scrapers because they often share their infrastructure publicly and lack sufficient funding and personnel to combat bots.[48] The open-source project Read the Docs

---

[41] Over 2,000 publishers use TollBit's network, including the Associated Press, Time, AdWeek, Forbes, and USA Today. *Frequently Asked Questions and Answers*, TollBit, https://tollbit.com/faqs/ (last visited Jan. 8, 2026).

[42] *AI Scraping Is On The Rise: TollBit State of the Bots - Q1 2025*, TollBit (June 11, 2025), https://tollbit.com/bots/25q1/.

[43] *Control Content Use for AI Training with Cloudflare's Managed Robots.txt*, Cloudflare Blog (July 1, 2025), https://blog.cloudflare.com/control-content-use-for-ai-training/.

[44] Reddit, Inc. v. Anthropic PBC, 3:25-cv-05643, (N.D. Cal.)

[45] Kyle Wiens (@kwiens), X (July 24, 2024, 8:15 AM), https://x.com/kwiens/status/1816128302542905620.

[46] *How Crawlers Impact the Operations of the Wikimedia Projects*, Wikimedia Diff (Apr. 1, 2025), https://diff.wikimedia.org/2025/04/01/how-crawlers-impact-the-operations-of-the-wikimedia-projects/.

[47] *AI Scraping Is On The Rise: TollBit State of the Bots - Q1 2025*, **TollBit** (June 11, 2025), https://tollbit.com/bots/25q1/.

[48] Niccolò Venerandi, *FOSS Infrastructure is Under Attack by AI Companies*, The Libre News (Mar. 20, 2025) https://thelibre.news/foss-infrastructure-is-under-attack-by-ai-companies/; Kyle Wiggers, *Open*

reported that "One crawler downloaded 73 TB of zipped HTML files in May 2024, with almost 10 TB in a single day."[49]

The scraping is also highly problematic for institutions that host content widely considered publicly beneficial. This includes galleries, libraries, archives, and museums (GLAMs). A joint initiative between the Centre for Science, Culture and the Law at the University of Exeter and the Engelberg Centre on Innovation Law & Policy at NYU Law called GLAM-E issued a report finding that of 43 respondents to their survey, "39 had experienced a recent increase in traffic. Twenty-seven of the thirty-nine respondents experiencing an increase in traffic attributed it to AI training data bots, with an additional seven believing that bots could be contributing to the traffic."[50]

The increase proved problematic for GLAMs in many ways, including that "Many respondents did not realize they were experiencing a growth in bot traffic until the traffic reached the point where it overwhelmed the service and knocked online collections offline."[51] The report also found that "Robots.txt is not currently an effective way to prevent bots from overwhelming collections."[52] One key conclusion they draw is that "The cultural institutions that host online collections are not resourced to continue adding more servers, deploying more sophisticated firewalls, and hiring more operations engineers in perpetuity."[53]

The Confederation of Open Access Repositories (COAR) reached similar conclusions. "Over 90% of survey respondents indicated their repository is encountering aggressive bots, usually more than once a week, and often leading to slowdowns and service outages."[54] COAR warns that "the recent rise in aggressive bot activity could potentially result in repositories limiting access to their resources for both human and machine users – leading to a situation where the value of the global repository network is substantially diminished."[55]

---

*Source Devs are Fighting AI Crawlers with Cleverness and Vengeance*, TechCrunch (Mar. 27, 2025), https://techcrunch.com/2025/03/27/open-source-devs-are-fighting-ai-crawlers-with-cleverness-and-vengeance/.

[49] Eric Holscher, *AI Crawlers Need to be More Respectful*, Read the Docs Blog (July 25, 2024), https://about.readthedocs.com/blog/2024/07/ai-crawlers-abuse/.

[50] Michael Weinberg, *Are AI Bots Knocking Cultural Heritage Offline?*, GLAM-E Lab (June 17, 2025), https://www.glamelab.org/products/are-ai-bots-knocking-cultural-heritage-offline/.

[51] Michael Weinberg, *Are AI Bots Knocking Cultural Heritage Offline?*, GLAM-E Lab (June 17, 2025), https://www.glamelab.org/products/are-ai-bots-knocking-cultural-heritage-offline/.

[52] Michael Weinberg, *Are AI Bots Knocking Cultural Heritage Offline?*, GLAM-E Lab (June 17, 2025), https://www.glamelab.org/products/are-ai-bots-knocking-cultural-heritage-offline/.

[53] Michael Weinberg, *Are AI Bots Knocking Cultural Heritage Offline?*, GLAM-E Lab (June 17, 2025), https://www.glamelab.org/products/are-ai-bots-knocking-cultural-heritage-offline/.

[54] Kathleen Shearer & Paul Walk, *Open Repositories Are Being Profoundly Impacted by AI Bots and Other Crawlers: Results of a COAR Survey*, COAR (June 3, 2025), https://coar-repositories.org/news-updates/open-repositories-are-being-profoundly-impacted-by-ai-bots-and-other-crawlers-results-of-a-coar-survey/.

[55] Kathleen Shearer & Paul Walk, *Open Repositories Are Being Profoundly Impacted by AI Bots and Other Crawlers: Results of a COAR Survey*, COAR (June 3, 2025), https://coar-repositories.org/news-updates/open-repositories-are-being-profoundly-impacted-by-ai-bots-and-other-crawlers-results-of-a-coar-survey/.

Sometimes, it is a nonprofit doing the scraping, as is the case with Common Crawl, an organization that claims it is for researchers but allows anyone to download its datasets for any reason and also actively assists AI companies in using them.[56] Common Crawl scrapes a sample of the internet every month, amounting to billions of webpages. This also makes Common Crawl a common source of data for AI training entities, most of which are for-profit and not research-focused. According to Alex Reisner in *The Atlantic*:

> [O}n many news websites, you can briefly see the full text of any article before your web browser executes the paywall code that checks whether you're a subscriber and hides the content if you're not. Common Crawl's scraper never executes that code, so it gets the full articles. Thus, by my estimate, the foundation's archives contain millions of articles from news organizations around the world, including *The Economist*, the *Los Angeles Times*, *The Wall Street Journal*, *The New York Times*, *The New Yorker*, *Harper's*, and *The Atlantic*.[57]

Moreover, Alex notes that while Common Crawl has told publishers it was removing publisher content from its archives, it has not modified any of its archives since 2016. Yet the search bar on Common Crawl returns "no captures" when users enter publisher websites.

## C. The Inadequacy of Robots.txt

The Robots Exclusion Protocol (robots.txt), a file that websites can use to direct bots on which web pages they can access, has been in existence for decades. The protocols' specifications were formally established by the Internet Engineering Task Force (IETF) in RFC 9309. The bots that are not listed in a robots.txt file are typically presumed to have access to webpages by default. Some may argue that robots.txt provides a sufficient technical measure to control AI bots, but this view is mistaken. Research indicates that bots frequently fail to comply with more restrictive robots.txt directives, and certain categories of bots, including AI search crawlers, often ignore robots.txt entirely.[58] Studies have also documented instances of malicious bots spoofing user agents to evade restrictions.[59]

First, an increasing number of websites are attempting to block bots through the use of robots.txt. TollBit reports that "the number of explicit disallow requests for AI bots (i.e., a single site disallowing a single bot would count as one) has increased from 559 to 2,165 (+287%). The average number of AI bots explicitly disallowed per website has grown from 2.2 to 8.6, a 4x increase."[60] While the most prominent GenAI bots are unsurprisingly blocked at higher rates, even the nonprofit Allen Institute for Artificial Intelligence's ai2bot-dolma is now blocked by 29% of sites on TollBit's network.

[56] *E.g.*, Pedro Ortiz Suárez, *Harnessing Common Crawl for AI and ML Applications*, YouTube (July 5, 2025), https://www.youtube.com/watch?v=jBY_oKuoz0o.

[57] Alex Reisner, *The Company Quietly Funneling Paywalled Articles to AI Developers*, The Atl. (Nov. 4, 2025), https://www.theatlantic.com/technology/2025/11/common-crawl-ai-training-data/684567/.

[58] Taein Kim, Karstan Bock, Claire Luo, Amanda Liswood, Chloe Poroslay & Emily Wenger, *Scrapers Selectively Respect robots.txt Directives: Evidence from a Large-Scale Empirical Study*, Proc. ACM Internet Measurement Conf. (IMC) (2025), https://doi.org/10.1145/3730567.3764471.

[59] Taein Kim, Karstan Bock, Claire Luo, Amanda Liswood, Chloe Poroslay & Emily Wenger, *Scrapers Selectively Respect robots.txt Directives: Evidence from a Large-Scale Empirical Study*, Proc. ACM Internet Measurement Conf. (IMC) (2025), https://doi.org/10.1145/3730567.3764471.

However, these blocking efforts have proven increasingly ineffectual.

> Publishers have attempted to block 4x more AI bots between January 2024 and January 2025 by disallowing them in their robots.txt file. However, AI bots are increasingly ignoring the robots.txt file. The percentage of AI Bot scrapes that bypassed robots.txt surged from 3.3% in Q4 2024 to 12.9% by the end of Q1 2025 (March). In March 2025, over 26 [million] scrapes from AI bots bypassed robots.txt for sites on TollBit.[61]

In Tollbit's Q2 report, they note that 13.26% of requests bypass robots.txt, indicating the trend is worsening. [62]The situation is further complicated by several GenAI companies that expressly state their intention to ignore robots.txt when scraping serves user prompts. This approach is adopted by Perplexity, Google, and Meta, among others.[63]

Moreover, robots.txt remains wildly underutilized. Among the top 10,000 domains on Cloudflare's extensive network,[64] only 37% maintained a robots.txt file at all.[65] The small percentage of websites with a robots.txt file likely indicates a general lack of awareness regarding robots.txt rather than deliberate permissiveness. Support for this interpretation can be found in the fact that 85% of websites using Cloudflare opted to block AI bots leading up to July 2024, when Cloudflare introduced a simple blocking option.[66] Cloudflare now reports that over one million customers block all AI scrapers.[67]

---

[60] *AI Scraping Is On The Rise: TollBit State of the Bots - Q1 2025*, TollBit (June 11, 2025), https://tollbit.com/bots/25q1/.

[61] *AI Scraping Is On The Rise: TollBit State of the Bots - Q1 2025*, TollBit (June 11, 2025), https://tollbit.com/bots/25q1/.

[62] TOLLBIT, *AI Scraping Is On The Rise: TollBit State of the Bots - Q2 2025*, (Sept. 8, 2025), https://tollbit.com/bots/25q2/.

[63] *AI Scraping Is On The Rise: TollBit State of the Bots - Q1 2025*, TollBit (June 11, 2025), https://tollbit.com/bots/25q1/.

[64] About a fifth of all internet traffic goes through Cloudflare. *Cloudflare, Inc.*, https://www.cloudflare.com/ (last visited Jan. 8, 2026).

[65] *Control Content Use for AI Training with Cloudflare's Managed Robots.txt*, Cloudflare Blog (July 1, 2025), https://blog.cloudflare.com/control-content-use-for-ai-training/.

[66] Alex Bocharov et al., *Declare your AIndependence: Block AI Bots, Scrapers and Crawlers with a Single Click*, Cloudflare Blog (July 3, 2024), https://blog.cloudflare.com/declaring-your-aindependence-block-ai-bots-scrapers-and-crawlers-with-a-single-click/.

[67] *Control Content Use for AI Training with Cloudflare's Managed Robots.txt*, Cloudflare Blog (July 1, 2025), https://blog.cloudflare.com/control-content-use-for-ai-training/. (These efforts appear to be working. The share of sites GPTBot crawls decreased almost 6.5% from the year before.)

Among sites that implement robots.txt, most block only the most prominent GenAI bots (OpenAI, Anthropic, etc.), and even these are blocked at low rates: GPTBot at 7.8%, Google-Extend at 5.6%, and other GenAI bots at less than 5%.[68] Given that there appears to be little rational basis for a site to block OpenAI while permitting Anthropic and Perplexity bots, this inconsistency is likely more attributable to a lack of awareness than to intentional policy.

Some entities, including Adobe[69] and llmstxt.org,[70] have proposed alternatives to robots.txt to provide website owners with more granular control over content access, but these solutions suffer from the same voluntary nature and compliance limitations as robots.txt.

## D. Referrals to Sources

What GenAI companies appear to overlook or flatly ignore is that bots do not serve the same economic or reputation-building function as humans visiting webpages. Indeed, the scraping by GenAI bots may not necessarily prove problematic in isolation. Some observers compare such scraping to indexing bots and question the distinction between them. However, TollBit's report indicates that such comparisons are weak at best, with "traffic from AI applications represented just 0.04% of all external referrals to sites in Q1 2025. This is insignificant when compared with Google, which delivers around 85% of traceable external visits."[71] A closer look at how often GenAI generates referrals reveals the lopsided benefits those companies enjoy at the expense of content creators.

### 1. Crawl-to-Referral Ratios

A revealing metric is the crawl-to-referral ratio exhibited by different bots. If GenAI entities directed humans to original publishers at rates similar to Google Search, then the argument for enforcing a strong website exclusion right would be weaker. However, TollBit found that "In Q1 2025, on average across TollBit sites, for every 11 crawls, Bing returns one human visitor to sites. This means that Bing's crawl-to-referral ratio is 11:1, up from 8:1 in Q4 2024 and 6:1 in Q3. Crawl-to-referral ratios for AI-only apps are as follows: OpenAI's ratio is 179:1, Perplexity's is 369:1, and Anthropic's ratio is 8692:1."[72] In September 2025, Tollbit reported that "100 crawls from Googlebot resulted in 454 referrals in Q2 2024, yet one year later, in Q2 2025, 100 crawls only resulted in 312 referrals."[73]

---

[68] *Control Content Use for AI Training with Cloudflare's Managed Robots.txt*, Cloudflare Blog (July 1, 2025), https://blog.cloudflare.com/control-content-use-for-ai-training/.

[69] Kyle Wiggers, *Adobe Wants to Create a Robots.txt-styled Indicator for Images Used in AI Training*, TechCrunch (Apr. 24, 2025), https://techcrunch.com/2025/04/24/adobe-wants-to-create-a-robots-txt-styled-indicator-for-images-used-in-ai-training/.

[70] *llms.txt: The /llms.txt file*, LLMs-txt (Sept. 3, 2024), https://llmstxt.org/.

[71] *AI Scraping Is On The Rise: TollBit State of the Bots - Q1 2025*, TollBit (June 11, 2025), https://tollbit.com/bots/25q1/. ('Other' platforms account for the remaining 15% and is comprised of search engines like Bing, Yahoo, and DuckDuckGo as well as social media platforms like LinkedIn and TikTok.")

[72] *AI Scraping Is On The Rise: TollBit State of the Bots - Q1 2025*, TollBit (June 11, 2025), https://tollbit.com/bots/25q1/.

[73] TOLLBIT, *AI Scraping Is On The Rise: TollBit State of the Bots - Q2 2025*, (Sept. 8, 2025), https://tollbit.com/bots/25q2/.

Cloudflare conducted a similar calculation.[74] As of August 4, 2025, Anthropic crawled 54,800 times for every referral it provided. OpenAI's ratio was 895:1, and Perplexity was 98:1. In contrast, Microsoft's was 45:1 and Google's was 4:1, reflecting their more traditional roles as traffic-generating search indexers.[75] According to Cloudflare's CEO, the problem has deteriorated over time. He stated in July 2025 that "OpenAI sent one visitor to a publisher for every 250 pages it crawled six months ago, while Anthropic sent one visitor for every 6,000 pages."[76]

In sharp contrast to GenAI entities, "Legacy search index crawlers would scan your content a couple of times, or less, for each visitor sent. A site's availability to crawlers made their revenue model more viable, not less."[77] Put differently, whereas traditional scraping generally represented a symbiotic relationship where websites sought to be scraped to facilitate discovery through search and ultimately reach human users, GenAI bots shatter this understanding of how the relationship should work.

Notably, the paucity of referrals is intentional. As Matteo Wong wrote for *The Atlantic*:

> Although ChatGPT and Perplexity and Google AI Overviews cite their sources with (small) footnotes or bars to click on, not clicking on those links is the entire point. OpenAI, in its announcement of its new search feature, wrote that "getting useful answers on the web can take a lot of effort. It often requires multiple searches and digging through links to find quality sources and the right information for you. Now, chat can get you to a better answer." Google's pitch is that its AI "will do the Googling for you." Perplexity's chief business officer told me this summer that "people don't come to Perplexity to consume journalism," and that the AI tool will provide less traffic than traditional search. For curious users, Perplexity suggests follow-up questions so that, instead of opening a footnote, you keep reading in Perplexity.[78]

Simply put, the scraping is designed to train GenAI systems with the express intent to supplant the need for people to visit webpages.

---

[74] *Control Content Use for AI Training with Cloudflare's Managed Robots.txt*, Cloudflare Blog (July 1, 2025), https://blog.cloudflare.com/control-content-use-for-ai-training/.

[75] *The crawl before the fall… of referrals: understanding AI's impact on content providers*, Cloudflare Radar AI Insights (July 1, 2025), https://radar.cloudflare.com/ai-insights.

[76] Mariella Moon, *Cloudflare CEO says people aren't checking AI chatbots' source links*, Engadget (June 20, 2025), https://www.engadget.com/ai/cloudflare-ceo-says-people-arent-checking-ai-chatbots-source-links-120016921.html.

[77] David Belson & Sam Rhea, *The Crawl Before the Fall… of Referrals: Understanding AI's Impact on Content Providers*, Cloudflare Blog (July 1, 2025), https://blog.cloudflare.com/ai-search-crawl-refer-ratio-on-radar/.

[78] Matteo Wong, *AI Search Engines Are Killing Curiosity*, The Atlantic (Nov. 25, 2024), https://www.theatlantic.com/technology/archive/2024/11/ai-search-engines-curiosity/680594/.

## 2. Google's AI Overview Impact

While Google's referral ratio ranks among the best, research from the Pew Research Center found that "About six-in-ten respondents (58%) conducted at least one Google search in March 2025 that produced an AI-generated summary."[79] Significantly, "Users who encountered an AI summary clicked on a traditional search result link in 8% of all visits. Those who did not encounter an AI summary clicked on a search result nearly twice as often (15% of visits)."[80] Even more telling, merely 1% of users clicked links within the AI Overviews.[81]

Additionally, users stopped browsing after 26% of searches when AI summaries were present compared to only 16% when summaries were absent.[82] Bain & Company corroborated these findings, reporting that 60% of searches end without users navigating to external websites when their AI summaries appear.[83] Other researchers estimate the impact at approximately 35%.[84]

A third study found that "On desktop, outbound clicks to external websites drop by about two-thirds. On mobile, the drop is nearly 50 percent. Many users see the [AI Overview, 'AIO'] as a complete answer. Clicking on links inside the AIO is rare—7.4 percent on desktop, 19 percent on mobile."[85] Moreover, "Even when users interacted with the AIO, they didn't go far. While 88 percent tapped 'Show more,' the median scroll depth was just 30 percent. Most (86 percent) only skimmed the content, and very few reached the bottom of the answer."[86]

---

[79] Athena Chapekis & Anna Lieb, *Google Users Are Less Likely to Click on Links When an AI Summary Appears in the Results*, Pew Research Center (July 22, 2025), https://www.pewresearch.org/short-reads/2025/07/22/google-users-are-less-likely-to-click-on-links-when-an-ai-summary-appears-in-the-results/.

[80] Athena Chapekis & Anna Lieb, *Google Users Are Less Likely to Click on Links When an AI Summary Appears in the Results*, Pew Research Center (July 22, 2025), https://www.pewresearch.org/short-reads/2025/07/22/google-users-are-less-likely-to-click-on-links-when-an-ai-summary-appears-in-the-results/.

[81] Athena Chapekis & Anna Lieb, *Google Users Are Less Likely to Click on Links When an AI Summary Appears in the Results*, Pew Research Center (July 22, 2025), https://www.pewresearch.org/short-reads/2025/07/22/google-users-are-less-likely-to-click-on-links-when-an-ai-summary-appears-in-the-results/.

[82] Athena Chapekis & Anna Lieb, *Google Users Are Less Likely to Click on Links When an AI Summary Appears in the Results*, Pew Research Center (July 22, 2025), https://www.pewresearch.org/short-reads/2025/07/22/google-users-are-less-likely-to-click-on-links-when-an-ai-summary-appears-in-the-results/.

[83] Natasha Sommerfeld et al., *Consumer Reliance on AI Search Results Signals New Era of Marketing*, Bain & Company (Feb. 19, 2025), https://www.bain.com/about/media-center/press-releases/20252/consumer-reliance-on-ai-search-results-signals-new-era-of-marketing--bain--company-about-80-of-search-users-rely-on-ai-summaries-at-least-40-of-the-time-on-traditional-search-engines-about-60-of-searches-now-end-without-the-user-progressing-to-a/

[84] Ryan Law & Xibeijia Guan, *AI Overviews Reduce Clicks by 34.5%*, Ahrefs Blog (Apr. 17, 2025), https://ahrefs.com/blog/ai-overviews-reduce-clicks/.

[85] Matthias Bastian, *Google's AI Answers are Changing User Behavior by Sharply Reducing Clicks to Websites*, The Decoder (Mar. 26, 2025), https://the-decoder.com/googles-ai-answers-are-changing-user-behavior-by-sharply-reducing-clicks-to-websites/.

[86] Matthias Bastian, *Google's AI Answers are Changing User Behavior by Sharply Reducing Clicks to Websites*, The Decoder (Mar. 26, 2025), https://the-decoder.com/googles-ai-answers-are-changing-user-behavior-by-sharply-reducing-clicks-to-websites/.

These findings may appear to conflict, but regardless of the specific percentage based on particular methodologies, the consistent trend of fewer people clicking links when presented with AI Overviews, which are trained on website content by scraping with the same bot Google uses to provide traditional search results, clearly disadvantages websites.

This trend also means that websites have limited alternatives to avoid a GenAI-mediated future.

### 3. Impact on Specific Websites

Another way to assess the issue is to examine how specific sites have fared since the launch of ChatGPT, which sparked the current wave of GenAI scraping. According to the *Wall Street Journal*, "Traffic from organic search to *HuffPost*'s desktop and mobile websites fell by just over half in the past three years, and by nearly that much at the *Washington Post*," and "Organic search traffic to [*Business Insider*'s] websites declined by 55% between April 2022 and April 2025."[87] "At the *New York Times*, the share of traffic coming from organic search to the paper's desktop and mobile websites slid to 36.5% in April 2025 from almost 44% three years earlier, according to Similarweb. The *Wall Street Journal*'s traffic from organic search was up in April compared with three years prior, Similarweb data show, though as a share of overall traffic it declined to 24% from 29%."[88]

The harm extends beyond news sites and major brands. Shira Ovide reported that one large sports betting website was subjected to thirteen million scraping attempts from AI bots in a month, resulting in only 600 human visitors. In contrast, Google's 15 million scraping operations of the same site generated *millions* of human visitors.[89]

At a more general level, Tollbit reports that some sites appear to be more sought after for scraping than others. For example, parenting content saw an AI request increase of 333% in Q2 of 2025. Deals and shopping content increased 111%. Human visits to lifestyle content decreased by 45% while AI requests to the same content increased 43%. For national news, human traffic decreased by 10%, while AI traffic increased by 65%.[90]

### 4. The Problem of Source Attribution

The presence of citations does not guarantee that the attribution is correct. Even if we (incorrectly) assume people click on links as often as they do with traditional search results, the citations prove more problematic than initially apparent. According to the Columbia Journalism Review's Tow Center for Digital Journalism:

> Even when these AI search tools cited sources, they often directed users to syndicated versions of content on platforms like Yahoo News rather than original publisher sites. This occurred even in cases where publishers had formal licensing agreements with AI companies.
>
> URL fabrication emerged as another significant problem. More than half of citations from Google's Gemini and Grok 3 led users to fabricated or broken URLs, resulting in error pages. Of 200 citations tested from Grok 3, 154 resulted in broken links.
>
> These issues create significant tension for publishers, who face difficult choices. Blocking AI crawlers might lead to loss of attribution entirely, while permitting them allows widespread reuse without driving traffic back to publishers' own websites.[91]

---

[87] Isabella Simonetti and Katherine Blunt, *News Sites Are Getting Crushed by Google's New AI Tools, Wall St. J. (June 10, 2025),* https://www.wsj.com/tech/ai/google-ai-news-publishers-7e687141

[88] Isabella Simonetti and Katherine Blunt, *News Sites Are Getting Crushed by Google's New AI Tools, Wall St. J. (June 10, 2025),* https://www.wsj.com/tech/ai/google-ai-news-publishers-7e687141

[89] Shria Ovide, How AI Bots Are Threatening Your Favorite Websites, Wash. Post, https://www.washingtonpost.com/technology/2025/07/01/ai-crawlers-reddit-wikipedia-fight/

[90] TOLLBIT, *AI Scraping Is On The Rise: TollBit State of the Bots - Q2 2025*, (Sept. 8, 2025), https://tollbit.com/bots/25q2/.

## E. Headless Browsers

Scraping bots are not the only issue websites must contend with. Fastly, an edge computing provider, reported that "In Q3 [2025] alone, we saw billions of requests from headless bots."[92] It's likely this approach to accessing content will increase.

There are three types of "headless" browsers, such as AI browsers. The first is masked web agents, in which infrastructure providers like Hyperbrowser and Browserbase let "an AI developer's agent" access websites at scale.[93] Next are third-party web agents, provided by companies such as Firecrawl, Apify, and Zyte, that crawl websites on behalf of an AI agent so the AI system doesn't access the website directly.[94] It's a way for AI companies to get around being blocked.

Finally, there are browser-driven web agents, such as Perplexity's Comet, OpenAI's Atlas, and Atlassian's Dia, which interact with websites directly, just as a human would.[95] These are consumer-facing products that claim to offer "agentic capabilities." The requests can even appear to come from the user's residential IP address. Because they appear to be human users using a traditional web browser, they are harder to block.

---

[91] Klaudia Jaźwińska & Aisvarya Chandrasekar, *AI Search Has a Citation Problem: We Compared Eight AI Search Engines. They're All Bad at Citing News*, Colum. Journalism Rev. (Mar. 6, 2025), https://www.cjr.org/tow_center/we-compared-eight-ai-search-engines-theyre-all-bad-at-citing-news.php.

[92] Fastly, Inc., *Fastly Threat Insights Report: From AI Crawlers to Headless Bots*, (Dec. 18, 2025), https://www.fastly.com/press/press-releases/fastly-threat-insights-report-reveals-organizations-are-fighting-back.

[93] TOLLBIT, *AI Scraping Is On The Rise: TollBit State of the Bots - Q2 2025*, (Sept. 8, 2025), https://tollbit.com/bots/25q2/.

[94] TOLLBIT, *AI Scraping Is On The Rise: TollBit State of the Bots - Q2 2025*, (Sept. 8, 2025), https://tollbit.com/bots/25q2/.

[95] TOLLBIT, *AI Scraping Is On The Rise: TollBit State of the Bots - Q2 2025*, (Sept. 8, 2025), https://tollbit.com/bots/25q2/.

A user may ask Atlas to gather information from sites that otherwise block OpenAI scraping bots, and the website may not recognize the AI browser as doing essentially the same thing (scraping content for RAG), but with a different method. Moreover, when AI browsers visit sites that use client-side overlays, where the text is covered by a paywall pop-up, it's as if the pop-up doesn't exist. The AI browser can still see the text.[96]

## F. Generative AI as Digital Gatekeepers

By positioning themselves between users and the websites they seek to visit, GenAI companies act as gatekeepers, demanding value before allowing users to access original content.

Research illustrates this phenomenon: one study found that for every 1,000 searches in the US using Google, only 360 lead users to the open web. The remaining 640 do not result in users clicking on any of the results Google surfaces (perhaps because of an AI Overview, snippet, or additional Google searches), and approximately one-third of all clicks direct users to Google-owned platforms such as YouTube, Google Flights, and Google Maps.[97]

Google could easily provide an option that allows websites to control whether their content is used to train GenAI models, such as Gemini and AI Overview summaries, while maintaining search visibility, but it chooses not to. Instead, the only real choice Google allows is to opt out of training the GenAI models. For summaries, Google presents a blunt, coercive, and binary choice: websites can either opt out of being scraped for both search results and summaries, or accept both. This means a site must choose between virtually disappearing from the Internet or allowing Google to scrape and use its content to generate AI Overviews that demonstrably reduce traffic.[98] As the complaint in *Chegg v. Google* puts it:

> This action challenges Google's abuse of its adjudicated monopoly in General Search Services to coerce online publishers like Chegg to supply content that Google republishes without permission in AI-generated answers that unfairly compete for the attention of users on the Internet in violation of the Antitrust laws of the United States. This conduct threatens to further entrench Google's generative search monopoly and to expand it into online publishing, restricting

---

[96] Aisvarya Chandrasekar and Klaudia Jazwinska, How AI Browsers Sneak Past Blockers and Paywalls, Colum. Journalism Rev. (Oct. 30, 2025), https://www.cjr.org/analysis/how-ai-browsers-sneak-past-blockers-and-paywalls.php.

[97] Rand Fishkin, *2024 Zero-Click Search Study: For every 1,000 EU Google Searches, only 374 clicks go to the Open Web; In the US, it's 360*, SparkToro Blog (July 1, 2024), https://sparktoro.com/blog/2024-zero-click-search-study-for-every-1000-us-google-searches-only-374-clicks-go-to-the-open-web-in-the-eu-its-360/.

[98] Charlotte Tobitt, *How Google Forced Publishers to Accept AI Scraping as Price of Appearing in Search*, Press Gazette (May 12, 2025), https://pressgazette.co.uk/platforms/how-google-forced-publishers-to-accept-ai-scraping-as-price-of-appearing-in-search/.

competition in those markets and reducing the production of original content for consumers.[99]

Paradoxically, as publisher traffic decreases, Google's advertising revenue has continued to grow, highlighting a disconnect between the content ecosystem on which Google depends and Google itself.[100] As the *Washington Post* notes, "When [Google] does show an AI answer on 'commercial' searches, it shows up below the row of advertisements. That could force websites to buy ads just to maintain their position at the top of search results."

Catherine Perloff wrote about this issue for *The Information*: "Publishers and advertisers have no other good option, even as the rise of ChatGPT has shown signs of sapping Google's power in the Web ecosystem. The phenomenon helps explain how Google is maintaining double-digit search ad growth–11.7% in the second quarter, slightly faster growth than the first quarter–despite signs that people are searching for information less on Google than they used to."[101] In other words, due to Google's advertising dominance, businesses are spending more just to remain in place.

Google's standard dodge when pressed about this problem is to claim the study in question used a flawed methodology[102] or stated the facts are, in fact, the opposite.[103] However, Google has not specified the alleged flaws, nor has it provided contradictory evidence.

## G. Impact on Content Creators

The distinction between a human visitor and a bot accessing the same site for ostensibly the same purpose is enormous. Even when humans are only visiting to find information, they may be exposed to advertisements that prompt them to click, subscribe to the site, or pay for premium content, and they may share information about the site with others, thereby helping to build the site's reputation. Bots fail to provide most of these benefits.

Bots neither view nor respond to advertisements; they don't subscribe to publishers, and while they may occasionally link to sources, they typically don't. In other words, while website owners create websites to connect with other humans for various reasons, bots satisfy none of these objectives. Bots primarily visit to strip-mine and exploit information while contributing virtually nothing to the website's sustainability. This is problematic because many site owners depend on traffic volume to generate sufficient ad impressions or other income to sustain their operations. This is also true for nonprofit websites that strive to make their content as open as possible. For instance, when a bot visits Wikipedia, it doesn't donate. Similarly, when a chatbot summarizes Wikipedia content, the human user doesn't see the request for donations.

---

[99] Chegg, Inc. v. Google LLC (1:25-cv-00543), District Court, District of Columbia

[100] Alphabet Inc., *Alphabet Announces Second Quarter 2025 Results*, Alphabet Investor Relations (July 23, 2025), https://abc.xyz/assets/cc/27/3ada14014efbadd7a58472f1f3f4/2025q2-alphabet-earnings-release.pdf.

[101] Catherine Perloff, *Advertisers Can't Quit Google, Despite Complaints About Traffic and Ads*, The Information (Aug. 5, 2025), https://www.theinformation.com/articles/advertisers-quit-google-despite-complaints-traffic-ads.

[102] *See, e.g.,* Catherine Perloff, *Advertisers Can't Quit Google, Despite Complaints About Traffic and Ads*, The Information (Aug. 5, 2025), https://www.theinformation.com/articles/advertisers-quit-google-despite-complaints-traffic-ads. ("Google says the study used a flawed methodology.")

[103] *See, e.g.,* Matteo Wong, The Great AI Training Heist, The Atlantic (June 24, 2025), https://www.theatlantic.com/technology/archive/2025/06/generative-ai-pirated-articles-books/683009/. ("Google argued that it was sending "higher-quality" traffic to publisher websites, meaning that users purportedly spend more time on the sites once they click over, but declined to offer any data in support of this claim.")

AI Overviews and similar AI products risk transforming publishers into content creators who are systematically cut off from the economic benefits of their own work. The CEO of IAB Tech Lab, a company that creates open technical standards across the ad-supported digital economy, Anthony Katsur, described publishers as "the plankton of the digital media ecosystem,"[104] underscoring their vulnerability in this evolving landscape.

Beyond revenue loss, the sheer volume of AI scraping imposes tangible operational burdens. Data center operators must also contend with the ongoing increase in server load from GenAI scrapers, which necessitates expanded connectivity and results in higher costs for content hosts. This directly impacts the operational expenses and financial sustainability for publishers.

As explained above, research consistently indicates a significant reduction in click-through rates from search results when AI Overviews are present, indicating a shift away from the click economy that has historically underpinned web monetization through advertising. As users increasingly obtain answers directly from AI without visiting source websites, the traditional model of driving traffic to produce ad impressions or subscriptions faces severe disruption. This requires a re-evaluation of how content value is measured and compensated.

The rapid expansion of the AI scraping market highlights tremendous demand for training data to fuel AI development. This demand has historically relied, in part, on AI companies' assumption that publicly available web data should be freely accessible for training. The conflict extends beyond mere copyright infringement. If content is universally treated as freely available input for AI, it devalues the labor, expertise, and financial investment required to create original, high-quality content. This existential threat directly undermines the financial viability and long-term sustainability of content creation industries.

## H. Broader Implications

The prolific exploitation has downsides for GenAI companies as well. First, GenAI companies may find that their models are increasingly trained on AI-generated content. Because that content is itself a rearrangement of other content, the quality, depth, variety, and veracity of the training data may decline with each iteration. Even if GenAI companies can filter out most AI-generated content when creating training datasets by investigating the data more methodically, it is unclear whether a scraping bot can perform a similar high-quality scrub during an RAG search.

---

[104] Joanna Gerber, Behind the IAB Tech Lab's New Initiative to Deal with AI Scraping and Publisher Revenue Loss, Ad Exchanger, https://www.adexchanger.com/publishers/behind-the-iab-tech-labs-new-initiative-to-deal-with-ai-scraping-and-publisher-revenue-loss/

Additionally, by reducing publisher funding, GenAI companies undermine those publishers' ability to continue operating or to maintain their current quality and breadth. This will result in less high-quality content for GenAI to train on.

Finally, society suffers from these developments. As people increasingly rely on GenAI for information while the quality of information sources declines, collective knowledge suffers. Society also loses when people create less content or limit public sharing in response to exploitation. The impact may be long-lasting. We cannot instantly restore websites that atrophy or shutter under the crush of GenAI. It takes time to revitalize these sectors and realign incentives so that it's worthwhile for people to resume creating and sharing. It is easier to tear the web down than to build it up.

Having established the technical, economic, and societal ramifications of GenAI-driven scraping, the discussion now turns to the legal doctrines that can address these challenges. Part 2 outlines the legal framework necessary to restore meaningful exclusionary rights for website owners.

# Part 2: The Legal Argument

In Part 1, we explored property, websites, how web pages load, how scraping works, and how GenAI fits into the scraping picture. Part 2 provides a legal analysis that identifies the property rights attaching to specific parts of websites, argues for a robust right to exclude, and offers illustrative examples of when and how that right should be applied.

## VII. What Kind of Property is a Website

As discussed in Section III, websites consist of at least three distinct elements: the server, the address, and the content. Here, I will describe which types of property law apply to each component and explain why these distinctions matter for establishing exclusion rights.

### A. Web Servers

A web server is a physical piece of hardware, so it fits neatly into traditional notions of personal property, no different from a lamp, necklace, or ball. While real property typically carries the highest exclusionary power, few doubt that we have the right to keep others from accessing our physical belongings without our permission or against our express wishes.

Servers are less central for the relevant analysis, however, because most website owners do not own the server on which their website runs. Instead, website owners are tenants renting space on the server alongside many other website owners, much like a high-tech apartment complex.

Just as an apartment complex allows renters to exclude others from their apartments, website owners have the right to exclude bots from their websites.

## B. Domain Names

Recall that a domain name is the website's address. A domain name, in and of itself, is not automatically protected as a copyright, patent, or trademark. It doesn't represent a "creative work" (like copyrighted code or art) or an "invention." Instead, it's primarily an online address.

When you register a domain name, you don't "own" it the way you own a physical object. Instead, you acquire a contractual right to use that specific name for a defined period (typically 1-10 years) from a domain registrar. As long as you renew it, you maintain that exclusive right to use it.

### *Kremen v. Cohen*

Domain names are undeniably valuable commercial assets. They can be bought, sold, and leased, and are often central to a business's online identity and brand. This economic value makes them property in a practical and legal sense. The case *Kremen v. Cohen* (2003)[105] addressed whether there is a property right in domain names. The Ninth Circuit used a three-part test:

> "First, there must be an interest capable of precise definition; second, it must be capable of exclusive possession or control; and third, the putative owner must have established a legitimate claim to exclusivity."[106]

The court found that:

> "Domain names satisfy each criterion. Like a share of corporate stock or a plot of land, a domain name is a well-defined interest. Someone who registers a domain name decides where on the Internet those who invoke that particular name—whether by typing it into their web browsers, by following a hyperlink, or by other means—are sent. Ownership is exclusive in that the registrant alone makes that decision. Moreover, like other forms of property, domain names are valued, bought and sold, often for millions of dollars, and they are now even subject to in rem jurisdiction, see 15 U.S.C. § 1125(d)(2)."[107]

Moreover, the court found that "Registering a domain name is like staking a claim to a plot of land at the title office. It informs others that the domain name is registered to the registrant and not to anyone else. Many registrants also invest substantial time and money to develop and promote websites that depend on their domain names."[108]

---

[105] Kremen v. Cohen, 337 F.3d 1024 (9th Cir., 2003)
[106] Kremen v. Cohen, 337 F.3d 1024 (9th Cir., 2003)
[107] Kremen v. Cohen, 337 F.3d 1024 (9th Cir., 2003)
[108] Kremen v. Cohen, 337 F.3d 1024 (9th Cir., 2003)

The court's reasoning and common sense lead to only one conclusion about domain names: people have a right to exclude others from their domain name and, therefore, from the website that depends on it. The question is whether a domain name is the same as the website for the purposes of the law. The answer must be yes.

A website cannot both exist and not exist by pretending that legal property analysis can occur only once someone has gained access to the content of the site. The domain name is effectively the website for legal purposes. The value of a domain name lies in its role as the gateway to the content. If you transfer content, you don't lose the website.[109] If you transfer the domain name, you lose control of the website. They are thus separable, and the domain name serves as a more accurate representation of what constitutes a website than the content alone for legal purposes. Since the only step before accessing content that involves a user (including bots) is entering the domain name to communicate with the server, it makes the most sense to think of the domain name as the website's entryway, defining the property boundary. If you can get past it, you can access the content.

The *Kremen* court is also instructive as to the type of personal property at issue. Domain names are not intellectual property, such as copyrights, trade secrets, or patents—instead, the court compared them to shares in a company and to a plot of land. This distinction is important because, whereas one cannot control access under the Copyright Act, patent law, or trademark law, one may control access under other forms of property law. IP law controls the use of the IP but doesn't prohibit possession. In other words, while there is generally no right to exclude under IP law, there *is* a robust right to exclude under other property law.[110]

## The Type of Agreement

Another indication of the type of property a website falls under is that the website itself (not its content) is not provided under a copyright license. Rather, it's owned through a standard contractual arrangement. A license indicates that the information it covers is protected by copyright. You license copyrighted works. You lease personal property.

More specifically, you typically don't "license" a domain name the way you license software or a copyrighted image. While the underlying legal concept of granting permission to use may seem similar, the established practices and terminology in the domain name industry are buying, selling, and leasing.

When you "buy" an original domain name, you are actually registering it with a domain registrar for a set period. You acquire the exclusive right to use that domain name for that period, essentially leasing it from the domain registry through the registrar. This is the closest you can get to "owning" a domain name. As long as you continue to renew, you maintain this right. It is also possible to "sell" a domain name, which means transferring your exclusive right to use that domain name to another party. This is a common practice in the "domain aftermarket" for premium domain names.

[109] Though your website might lose some or all of its value.

[110] *E.g.,* I can't use IP to keep you from reading this paper, but I can use property law to keep you from taking a printed version of it from my hands.

Finally, there is the option of leasing an existing domain name. In a domain lease/rental agreement, the current registrant (the owner of the right to use) grants another party the right to use the domain name for a defined period, typically in exchange for recurring payments (e.g., monthly or annually). The original registrant retains control and registration of the domain name, while the lessee/renter is granted the right to use the domain (point it to their website, use it for email, etc.). The agreement typically outlines the terms of use, payment schedule, and what happens at the end of the term (e.g., the option to purchase or return it to the original owner).

## Domain Name Conclusion

While a property analogy for a website might break down under the misconception that websites always exist in a completed form and that the site's content is the website itself rather than merely part of it, a proper analysis clarifies the misunderstanding. Websites, unlike the content on them, are rivalrous: if I own a website, others cannot simultaneously own it. I cannot own Google.com while Google owns it. Additionally, website access is excludable because I can exclude others from accessing the website, just as I can exclude people from accessing the contents of my backpack.

The domain name is the appropriate level at which courts should examine property rights pertaining to access under the vast majority of circumstances.[111] This conclusion should feel intuitive. It would be odd to say that a website is not allowed to exclude people (i.e., it must allow everyone access to the website, including malicious IP addresses). The reason is that allowing malicious users onto a website could destroy its value (by damaging the server, deleting code, stealing personal data, ruining a brand’s reputation, or through any number of other harmful actions).[112] Likewise, allowing GenAI companies to copy content for their unilateral benefit while undermining the sustainability or general purpose of the scraped website as a matter of law, leaving only a post hoc claim of copyright infringement as the only practical way to defend one’s site, makes no sense.

Ultimately, the domain name is what people typically refer to when discussing a website, and it is the best way to think about website ownership technically and legally.[113] It is also the most practical level for website owners to implement technical safeguards to control access to their property. It would make less sense for a website owner to have to accept whatever safeguards someone higher or lower in the tech stack happens to use,[114] rather than for the website owner to choose the technical measures best suited for their website. In short, website owners should be the ultimate deciders of who they allow on their property, and the owner of the domain name is the best indicator of who owns the website.[115]

---

[111] As with all law, there will be edge cases. It’s possible to have a website without a domain name, for example. However, most websites, including those most interesting to GenAI scrapers, have domain names.

[112] Forcing websites to allow everyone to visit the site may well evoke a compelled speech argument.

[113] *E.g.,* Wikipedia also identifies the domain name as a key element of a website: “A website (also written as a web site) is any web page *whose content is identified by a common domain name* and is published on at least one web server. Websites are typically dedicated to a particular topic or purpose, such as news, education, commerce, entertainment, or social media.” (emphasis added) "Website," *Wikipedia*, https://en.wikipedia.org/wiki/Website (last visited Jan. 9, 2026).

It is this exclusive contractual right to control the digital space at that name's address that satisfies the three-part property test, making it the focal point of the property, not the content. The domain name is the most appropriate layer for exercising the paramount property right—the right to exclude—making it more than "just a name" in a legal context.

## Content

A website is made up of code, just as books are made up of words. The underlying code, including HTML, CSS, JavaScript, and any server-side programming code (e.g., Python, PHP, Ruby) that makes the website function, is considered a "literary work" and is automatically protected by copyright as soon as it is created and fixed in a tangible medium. The code is not the website. Code is necessary for a website to exist as personal property, but it is not sufficient on its own.

As with code, the visual elements, graphical user interface (GUI), and overall arrangement can be protected by copyright if they are original and creative. These elements are necessary components of a website, but standing alone they do not constitute the website as personal property. The website, as a unified digital asset accessible through its domain name and existing as an excludable, rivalrous resource, transcends any single part.

It is worth noting that website owners can choose to allow the individual content owners who post on the website to control access to the content owners' data. For example, the Substack website belongs to Substack, but it gives the people who write blogs on their platform some control over access by allowing paywalls.[116] Similarly, one's Instagram account lives on Instagram's (Meta's) website. The account owner does not have a separate website; instead, they have a unique URL on Instagram's website. This layered structure means that exclusion rights may be distributed between the platform and the user, and that legal remedies for unauthorized scraping may depend on the specific contractual and technical arrangements in place.

---

[114] The tech stack in this case refers to the parties that provide the infrastructure that makes websites possible. Just as it would make little sense for the federal government to control who can access your house, it makes little sense to declare ICANN the best place to evaluate property ownership. To build out the analogy: ICANN (Internet Corporation for Assigned Names and Numbers) would be the federal government. Domain Registries (e.g., Verisign for .com, PIR for .org) are the state or provincial governments. Domain Registrars (e.g., GoDaddy, Namecheap) are the county or municipal governments. Domain Name Owners are the property owners. Content Owners are the tenants or leaseholders. Other elements, such as DNS servers, are outside the scope. DNS servers would fit into the analogy as the public transportation system and the official directories.

[115] This even applies to the fediverse. For example, the Mastodon user controls the terms of the license for their content, but the platform owner controls the access to the underlying property (the website). If a bot bypasses the platform's technical access controls to access your "followers-only" content, the platform owner still has a viable claim of trespass to chattels against the bot for unauthorized access to the website's server/property (the chattel). Moreover, the "impairment as to its value" clause in Restatement § 218(b) remains applicable to the platform owner in this context. If the platform's value lies in its promise of a controlled, non-exploitative environment for users, yet a GenAI scraper can easily bypass privacy settings to harvest content, the platform loses its core competitive advantage, and its business model is undermined. The resulting loss of user trust, reputation, and potential financial contributions (subscriptions, donations) directly impairs the platform's value as a commercial asset.

## Conclusion

Three primary elements make up a website. All of them can be property, and more than one can control access, but domain names are best positioned, for practical and legal reasons, for an analysis of exclusion rights for websites as a whole.

In most instances, without the domain name, a bot cannot find the server's IP address to connect with a website. Thus, the domain name is the legal "front door," and the owner's primary means of imposing an exclusionary rule before the server even begins processing the request.[117]

Moreover, the website owner's decision to implement technical barriers (such as WAFs or IP blocking) is enforced at the domain name level through the connection to the server. The domain name, therefore, is the point of intervention where the right to exclude is exercised.

With the property status of websites established, the focus now shifts to the practical and doctrinal centerpiece of property law: the right to exclude and its enforcement in the digital context.

# VIII. The Right to Exclude

Exclusion rights without enforcement mechanisms are meaningless.[118] Yet courts remain reluctant to enforce website owners' rights even against aggressive GenAI scraping.

---

[116] If Substack does not implement sufficient technical measures to block scrapers, then those scrapers have access to the website. However, if the scrapers bypass the paywalls for the individual content, that would be a trespass to that particular blog.

[117] It's possible for multiple URLs to point to the same content. But while the URL's path or query parameters may change (/recipe.html vs. /recipes?lasagna=true), the domain name (e.g., foodblogger.com) remains constant. This fact proves the chapter's central theory: The content (the lasagna recipe) may be non-rivalrous and accessible via different paths (URLs), but the personal property (the domain name and website architecture) remains a single, excludable, rivalrous asset. The domain name, not the specific content file, is the consistent boundary that the trespasser must cross.

[118] While *Jacque v. Steenberg Homes Inc.* was about real property, the logic is instructive: "Society has an interest in punishing and deterring intentional trespassers beyond that of protecting the interests of the individual landowner. Society has an interest in preserving the integrity of the legal system. Private landowners should feel confident that wrongdoers who trespass upon their land will be appropriately punished."

When plaintiffs bring claims against GenAI companies, they typically skip past even attempting trespass to chattels because the law has been so neutered over the last twenty years.[119] When protecting access is mentioned in lawsuits or scholarly work, the focus tends to be on the Computer Fraud and Abuse Act (CFAA).[120] However, the CFAA often proves similarly ineffective because, unlike with other property, website content is presumed to be fully public if it is public at all, and courts have been reluctant to support claims that scraping such "publicly available" content should be a crime, even when the scrapers knowingly and intentionally circumvent technical measures to acquire the data.

Claims of conversion are similarly ineffectual because the scraped content is protected by copyright. In most cases, copying copyrighted material will not deprive the copyright owner of their rights to a significant extent, thereby not triggering conversion. Privacy and unjust enrichment claims have similarly run into dead ends in GenAI litigation.

Ultimately, scrapees are generally left with two claims: copyright infringement (assuming they own the copyright to the content) and breach of contract. Breach of contract is typically based on browsewrap (hyperlinks to terms of service at the bottom of websites that few people ever click, and which are therefore usually not enforceable because there is no notice), with the breach being the scraping of the site in violation of those terms.[121]

Courts have been willing to allow breach-of-contract claims to enforce website owners' rights. For instance, in *hiQ v. LinkedIn*, LinkedIn asserted that hiQ violated the CFAA. The court held that the CFAA did not apply because the content at issue was public, but it agreed that hiQ violated LinkedIn's terms of service, which prohibited scraping. Therefore, LinkedIn prevailed.

However, even if terms of service are enforceable, such claims pertaining to GenAI are often preempted by copyright law because the GenAI bots scrape copyrighted content, and copyright preemption sweeps breach-of-contract claims aside.[122] This means that the only claims based on scraping that tend to survive are those based on copyright law. Unfortunately for website owners, copyright law allows for the affirmative defense of fair use. If the scrapers win the fair use argument, as Meta and Anthropic did at the lower courts, they essentially win on everything.[123]

---

[119] When I compiled a tally of all initial claims in over 80 GenAI lawsuits, the claims of trespass to chattels did not appear once.

[120] *See*, e.g., Ben Sobel's overview in *A New Common Law of Web Scraping,* Lewis & Clark Law Review, which notes the trend of cases focusing on the CFAA while similarly giving little attention to trespass to chattels.

[121] I advocate for terms that prohibit scraping, though you can only see the terms if you are able to see the content. Terms are more of a belt-and-suspenders approach.

[122] "On and after January 1, 1978, all legal or equitable rights that are equivalent to any of the exclusive rights within the general scope of copyright as specified by section 106 in works of authorship that are fixed in a tangible medium of expression and come within the subject matter of copyright as specified by sections 102 and 103, whether created before or after that date and whether published or unpublished, are governed exclusively by this title. Thereafter, no person is entitled to any such right or equivalent right in any such work under the common law or statutes of any State." 17 U.S. Code § 301.

[123] Anthropic's use of pirated material to create a library was not deemed fair use, but their scraping of the content to train models was.

Moreover, if a site doesn't own the copyright in the content on its site, as is the case with Reddit, X, Bluesky, and other social media platforms, then the website cannot assert a copyright claim.[124] In effect, sites can be largely defenseless against scrapers unless they make themselves difficult to find online by blocking all bots (including search indexers like Google, which also use content for GenAI) or moving their content behind account logins,[125] which isn't appetizing to smaller sites that need to be visible to encourage traffic.

Fortunately, the claim of trespass to chattels is ripe for revitalization. The question of what the terms of service mentioned above govern is instructive. The terms at issue on LinkedIn pertained to the LinkedIn website. By giving force to those terms, the court agreed that the website owners had a property right in the website, allowing them to create those terms. If the website were not property, the website owner would not be allowed to create enforceable terms about the website. This makes sense. I cannot make up contractual terms for how you use your personal computer, but you can, for example. We can only create contractual terms over property you control. In other words, the court has implicitly recognized that websites are property, thereby implying that website owners have the right to exclude others.

## A. Trespass to Chattels

Trespass to chattels is a common law tort designed to protect an individual's possessory interest in personal property, known as chattels. Under the Restatement (Second) of Torts Sec. 217 (1965), "A trespass to a chattel may be committed by intentionally (a) dispossessing another of the chattel, or (b) using or intermeddling with a chattel in the possession of another."[126] Notably, the proposal of this paper does not require a change to the Restatement to give the force of law to trespass to chattels.

Furthermore, Sec. 218 states that:

> One who commits a trespass to a chattel is subject to liability to the possessor of the chattel if, but only if, (a) he dispossesses the other of the chattel, or (b) the chattel is impaired as to its condition, quality, *or value*, or (c) the possessor is deprived of the use of the chattel for a substantial time, or (d) bodily harm is caused to the possessor, or harm is caused to some person or thing in which the possessor has a legally protected interest. (emphasis added)

Two comments, e and h, are also relevant to this paper's analysis. Comment e states:

---

[124] *See, e.g.,* X Corp. v. Bright Data Ltd., No. 3:23-cv-03698-WHA (N.D. Cal. May 9, 2024) (Granting X Corp. the ability to restrict others from using user-owned X content would exceed its rights under the Copyright Act, as it holds only a non-exclusive license to that content.)

[125] The reason this may be successful is because scrapers subverting such measures to acquire non-publicly available information may allow for a viable CFAA claim.

[126] Restatement (Second) of Torts § 217 (Am. Law Inst. 1965).

> The interest of a possessor of a chattel in its inviolability, unlike the similar interest of a possessor of land, is not given legal protection by an action for nominal damages for harmless intermeddlings with the chattel. In order that an actor who interferes with another's chattel may be liable, his conduct must affect some other and more important interest of the possessor. Therefore, one who intentionally intermeddles with another's chattel is subject to liability only if his intermeddling is harmful to the possessor's materially valuable interest in the physical condition, quality, *or value of the chattel*, or if the possessor is deprived of the use of the chattel for a substantial time, or some other legally protected interest of the possessor is affected as stated in Clause (c). Sufficient legal protection of the possessor's interest in the mere inviolability of his chattel is afforded by his privilege to use reasonable force to protect his possession against *even harmless interference*. (emphasis added)

Sec. 218 comment h says:

> An unprivileged use or other intermeddling with a chattel which results in actual impairment of its physical condition, quality *or value to the possessor makes the actor liable for the loss thus caused*. In the great majority of cases, the actor's intermeddling with the chattel impairs the value of it to the possessor, as distinguished from the mere affront to his dignity as possessor, only by some impairment of the physical condition of the chattel. There may, however, be situations in which *the value to the owner of a particular type of chattel may be impaired by dealing with it in a manner that does not affect its physical condition.... In such a case, the intermeddling is actionable even though the physical condition of the chattel is not impaired*. (emphasis added)

This paper concerns the harm to the website's value.

## B. Court Interpretation

There are two ways to examine how trespass to chattels is interpreted and applied under the law: how courts treat such claims and how scholars view them. I will begin with an overview of how courts have handled trespass claims over the years.

### 1. Judicial Treatment of Scraping

Case law often oversimplifies the mechanics of scraping; yet the detailed process and terminology are important to reaching the correct legal outcome in scraping cases. Notably, court opinions rarely describe how scraping occurs. Instead, courts treat websites as binary entities: they're either there or they're not. This oversimplified conception is evident in numerous scraping cases, even as courts acknowledge some preventive blocking measures.

For example, in *Craigslist, Inc v 3Taps*, the court summarizes scraping in a single sentence: “Craigslist alleges that 3Taps copies (or “scrapes”) all content posted to Craigslist in real time, directly from the Craigslist website.”[127] In *Bright Data, Inc. et al. v. Meta Platforms*, et al. the court states that “Under the scraping umbrella, there exist two variations: “logged-in” scraping, which involves scraping data that is behind a password-protected website, and “logged-out” scraping, which involves scraping data that is viewable without a password, but may still be subject to restrictions on access, use, and rate and data limits.”[128] *Bright Data* appears to assume that website content is available in its entirety at all times, with the only variable being whether a user must be logged in to view it. In *hiQ Labs v LinkedIn Corporation*, the description is again reduced to a single sentence: “Using automated bots, [hiQ] scrapes information that LinkedIn users have included on public LinkedIn profiles, including name, job title, work history, and skills.”[129]

These cases and similar factual descriptions in others are likely shaped by the specific legal arguments plaintiffs have advanced. Perhaps courts focus on events after bots have gained access because that timing aligns with plaintiffs' theories of liability. However, this paper aims to illuminate a potentially more promising legal path for future plaintiffs.

While the cases above were primarily or entirely about CFAA claims, the right to exclude pairs best with trespass-to-chattels claims in most instances. The simple reason is that a violation of the right to exclude is a trespass. Unfortunately, courts typically disfavor trespass-to-chattels claims in scraping cases. However, courts have not declared that websites are not personal property; rather, they have found the interference in previous cases insufficiently substantial to warrant a trespass claim. Courts did not always favor those who extract data from others. Initially, courts adopted a broad interpretation, allowing website owners to assert control over their digital infrastructure against unwanted automated access.

The advent of the internet and electronic communications presented a novel challenge for the application of this historically physical tort. Courts began extending the concept of trespass to chattels to address digital interference, reasoning that "electrical signals traveling across networks and through proprietary servers may constitute the contact necessary to support a trespass claim."[130] This marked a significant adaptation of common law principles to a new technological frontier and underscored the adaptive capacity of common law, allowing courts to interpret existing torts in ways that could encompass new forms of property interference arising from technological advancements.

The evolution of trespass to chattels in the context of web scraping is perhaps best understood by reviewing key judicial decisions that have shaped its interpretation and applicability. Three cases from the first decade of the World Wide Web illustrate the rapid erosion of the right to exclude by the courts: CompuServe Inc. v. Cyber Promotions, Inc. (1997); eBay Inc. v. Bidder's Edge, Inc. (2000); and *Intel Corp. v. Hamidi* (2003).[131]

---

[127] Craigslist, Inc v. 3Taps, Inc et al, No. 3:2012cv03816 - Document 101 (N.D. Cal. 2013)
[128] Bright Data, Inc. et al. v. Meta Platforms, et al.
https://law.justia.com/cases/delaware/superior-court/2023/n23c-01-229-skr-ccld.html
[129] hiQ Labs, Inc. v. LinkedIn Corporation, No. 17-16783, D.C. No. 3:17-cv-03301-EMC, https://cdn.ca9.uscourts.gov/datastore/opinions/2022/04/18/17-16783.pdf
[130] 94 Cal. App. 4th 336

## **CompuServe Inc. v. Cyber Promotions, Inc.**

*CompuServe Inc. v. Cyber Promotions, Inc*. is a pioneering case applying trespass to chattels to digital activities. CompuServe, a prominent online service provider, sued Cyber Promotions, which sent a substantial volume of unsolicited email advertisements, or spam, to CompuServe's subscribers. CompuServe argued that this mass mailing imposed a significant burden on its proprietary computer equipment, which had finite processing and storage capacity, thereby diminishing its value and utility. Despite repeated demands to cease and desist, Cyber Promotions continued its activities, even falsifying point-of-origin information in the email headers to conceal the true source of the messages.

The U.S. District Court for the Southern District of Ohio concluded that Cyber Promotions' transmission of unsolicited emails constituted a trespass to CompuServe's chattels. The court then granted CompuServe's request for a preliminary injunction, effectively preventing Cyber Promotions from sending further unsolicited electronic messages. The court's determination rested on several foundational points. It recognized CompuServe's possessory interest in its computer systems and found that Cyber Promotions' intentional, unauthorized use of these systems by directing emails to CompuServe's equipment constituted actionable interference. The court emphasized that the high volume of unsolicited messages placed a "significant burden" on CompuServe's equipment, impairing its condition and value, thereby satisfying the damage requirement for trespass to chattels.

Furthermore, the court rejected Cyber Promotions' First Amendment defense, affirming CompuServe's right as a private entity to control access to its systems. The ruling was groundbreaking, establishing that electronic signals could be considered “sufficiently physically tangible"[132] to support a trespass cause of action, thereby bridging the gap between traditional physical torts and digital phenomena. Notably, relying on Sec. 218(b) of the Restatement, the court found that dispossession of property was not necessary for a successful claim. Instead, the court stated that “It is clear from a reading of Restatement § 218 that an interference or intermeddling that does not fit the § 221 definition of 'dispossession' can nonetheless result in defendants' liability for trespass.”[133]

The court's finding that a "significant burden" constituted sufficient damage reflected the technological limitations prevalent in the late 1990s, when even large volumes of email could genuinely impact server performance and incur costs, making them readily quantifiable as "impairment" or "diminution of value".

[131] 30 Cal. 4th 1342 (2003)
[132] eBay, Inc. v. Bidder's Edge, Inc., 100 F. Supp. 2d 1058 (N.D. Cal. 2000)
[133] eBay, Inc. v. Bidder's Edge, Inc., 100 F. Supp. 2d 1058 (N.D. Cal. 2000)

## eBay, Inc. v. Bidder's Edge, Inc.

Following *CompuServe*, *eBay Inc. v. Bidder's Edge*, Inc. further extended the application of trespass to chattels to the nascent field of web scraping. eBay, a prominent online auction platform, sought a preliminary injunction to prevent Bidder's Edge (BE), an auction data aggregator, from using automated bots to scrape data from eBay's website. BE's activities violated eBay's terms of use, and BE continued collecting data despite repeated requests from eBay to cease. eBay asserted several causes of action, including trespass to chattels, to halt BE's operations.

The District Court for the Northern District of California granted eBay's motion for a preliminary injunction, concluding that eBay had shown a sufficient likelihood of success on its trespass to chattels claim. The injunction barred BE from using any automated query program, robot, or similar device to access eBay's computer systems or networks to copy any part of eBay's auction database. The court found BE's use of crawlers intentional and unauthorized, given BE's violation of eBay's terms of use and disregard for eBay's requests to stop. Although the court acknowledged that BE's interference might not have been "substantial" in terms of severe system disruption, it nonetheless held that "any intermeddling with or use of another's personal property" could establish possessory interference. The court reasoned that upholding personal and intellectual property rights was essential to the proper functioning of the internet. Notably, the court expressed respectful disagreement with other district courts that had found that "mere use of a spider to enter a publicly available website to gather information, without more, is sufficient to fulfill the harm requirement for trespass to chattels."[134] Despite this nuance, the court found a sufficient likelihood of harm to justify the injunction.

## Intel Corp v. Hamidi

*Intel Corp. v. Hamidi* marked a pivotal moment, significantly narrowing the application of trespass to chattels in the digital realm. Kourosh Kenneth Hamidi, a former Intel Corporation employee, sent a series of mass emails over two years, criticizing the company's employment practices to approximately 35,000 current employees via Intel's internal email system. Hamidi did not breach any security measures to access these email addresses and offered recipients the option to opt out. The emails caused no physical damage or functional disruption to Intel's computer systems. Intel, however, claimed that the emails distracted employees and reduced productivity, and accordingly filed a trespass-to-chattels lawsuit seeking an injunction to prevent further emails.

The Supreme Court of California held that sending unsolicited emails that neither physically damage nor functionally impair a computer system does not constitute an actionable trespass to chattels under California law. The court reversed the lower courts' rulings granting Intel the injunction. The court's decision hinged on a strict interpretation of the "damage" requirement for trespass to chattels. It articulated that the tort requires an actual impairment to the condition, quality, or value of the chattel, or its dispossession. The court explicitly stated that "mere unwanted electronic communication without such harm is insufficient" and that Intel didn't show

[134] eBay, Inc. v. Bidder's Edge, Inc., 100 F. Supp. 2d 1058 (N.D. Cal. 2000)

that Hamidi caused “some measurable loss from the use of its computer system.”[135] Rather than overrule *CompuServe*, the court distinguished Hamidi's actions from *CompuServe*, where the sheer volume of communications demonstrably burdened the system. Intel's assertion of lost employee productivity was characterized as harm to its business interests, rather than as an injury to its property itself, and the court clarified that the tort was not designed to address "impairment by content." This ruling profoundly altered the landscape, moving away from the broader "any intermeddling" standard suggested in eBay and establishing a significantly higher bar for plaintiffs to prove actionable harm in digital trespass cases. The distinction between harm to the chattel and economic or business harm became a critical impediment for website owners seeking to use this tort against non-disruptive scraping.[136]

**<u>Summary</u>**

The shadow of *Intel Corp. v. Hamidi* has long chilled digital trespass claims by requiring proof of 'measurable loss' to computer functionality. However, the *Hamidi* court focused on server impairment only because it failed to consider the website's independent value as a chattel. Modern GenAI scraping differs from the spam at issue in Hamidi. It does not merely use the system; it extracts the core economic value of the property, causing a totalizing harm that threatens the digital asset's viability. We must move beyond the broken server requirement to recognize that destroying a website’s business model through unauthorized access is a cognizable injury to property.

Scholarly analysis reveals additional limitations in current approaches.

## C. Scholar Interpretation

The scholarship on trespass to chattels in the context of the internet is relatively thin. Instead, most scholars have focused on criminal trespass, particularly the Computer Fraud and Abuse Act (CFAA). To be clear, this paper is not about the CFAA or criminal conduct; it’s about property and tort law, so the comparison is not 1:1. However, the framework used to analyze CFAA claims seems to have seeped into tort claims, and it is worthwhile to note the differences and where the analysis goes awry.

Perhaps a representative paper on the CFAA is Orin Kerr’s *Norms of Computer Trespass*.[137] Notably, in the title and multiple instances throughout the paper, *Norms* does not specify criminal trespass when referring to trespass, although that is the only form of trespass Kerr’s paper focuses on. *Norms* argues that unless someone circumvents an authentication requirement, such as a login screen or accessing a website without an account when an account is otherwise required, the access is not criminal. In general, I agree with his paper. For instance, he correctly notes that courts "must identify the best rules to apply from a policy perspective, given the state of technology and its prevailing uses."

---

[135] Intel Corp. v. Hamidi, 71 P.3d 296, 306-307 (Cal. 2003).

[136] It’s notable how close the decision in *Hamidi* was. The district court, appellate court, and three of the seven judges on the California Supreme Court ruled in favor of Intel.

[137] To be fair, the paper was published in 2016, before the transformer architecture underlying all major GenAI platforms was even invented. His argument may have made more sense for tort trespass then, before rampant, abusive, and exploitative bots became far more common.

*Norms*' framework for analyzing criminal trespass is also helpful. I agree, for example, that the means of access are important. "An open window isn't an invitation to jump through the window and go inside. If there's an open chimney or mail drop, that's not an invitation to try to enter the store. Barring explicit permission from the store owner, the only means of permitted access to a commercial store is the front door."[138]

As I argue below, the means of entry matter. If a scraper must intentionally circumvent technical measures to access a website (i.e., access can't occur through normal use or negligence), that is sufficient to find the means inappropriate. This must certainly be true when the only purpose of the circumvention is to extract from and exploit the website.

But the *Norms* paper and I sharply differ on some of the underlying logic leading to that position. In short, I agree with Kerr's conclusion because I believe that criminal conduct should be held to a higher standard than civil claims, not necessarily because of his analogies or policy arguments. To clarify how our arguments differ, I will make four primary critiques of *Norms*: (i) analogizing to real property is the wrong analogy, (ii) the line between what is private and public is blurrier than *Norms* implies, (iii) the law does not limit people to protecting their property only after the intrusion, and (iv) the Internet is not a monolith.

### 1. Analogizing to Real Property

The first critique is that *Norms* draws comparisons from real property rather than personal property. It uses the phrase "physical-world trespass," but it is always in reference to real property. This is an odd move because websites are not real property. They are not land, a house, or a building. The more natural comparison would be to personal property, specifically to personal property that is an opaque container, like a backpack, chest, or box. The reason is that websites, like a box, can contain content, and, like a box, people cannot see the content until it is accessed.

Also, importantly, like a box, the owner controls who can look inside. In other words, they control access. *Norms*, instead, declares that if you "set up shop at a public fair,"[139] you lose all rights to exclude anyone from your space, which necessarily includes known thieves. It seems that, to his mind, public fairs and the Internet are places of clear-cut absolutes regarding property rights, where one must either entirely relinquish the right to exclude or exclude everyone who doesn't use a particular method of exclusive entry.

I don't want to quibble over analogies, however. It's not even clear to me that an analogy is necessary when working from the plain meaning of the Restatement's text. But if we are to use any, it makes sense to use the correct category of comparison.

---

[138] *Norms* at 1152
[139] *Norms* at 1163

## 2. Private v. Public

*Norms* also makes sweeping generalizations about what is public and private. In Kerr's view, if it is on the internet and not hidden behind a login, it is public and it must be freely accessible to anyone.[140] As he puts it, "Everyone can visit a public website."[141] He further claims that "Publishing on the Web means publishing to all." This, of course, is not how personal property works. What *Norms* calls "public" has always been subject to caveats. Websites have never been obligated to allow known criminals or other malicious actors to access their websites, for example.

As another example, *Norms* seems to argue that news sites like the *Wall Street Journal* and the *New York Times* should not be allowed to share any stories publicly and freely as part of a "get three free articles" type of paywall without allowing anyone to access *all* articles for free because if anyone can choose any articles as their three free articles, that must mean no articles are "private information." The self-evident test, in this view, is that the articles are not kept under lock and key at all times. In other words, in order to have a property right in something online, you would have to hide everything.

A site's content must not be presumed public in the sense of a public park or road just because some people are allowed to view it without restriction, any more than a backpack's contents are public just because an owner allows some favored friends and family to go through it. Being publicly available is not the same as being common property. Allowing some or most people to access something does not require everyone to have such access.

## 3. Proactive v. Reactive

When discussing exclusion, *Norms* exclusively focuses on the right of real property owners to react to trespassers unless the door of their property is locked.[142] Similarly, when speaking of a house, *Norms* declares that "If the owner grants you permission but later revokes it, your authorization expires with the revocation."[143] While this may be true, it is also true that if the owner *never* grants you permission to enter, even if the store or house door is unlocked, you are similarly without authorization.

*Norms'* stance overlooks the important right of property owners (both real and personal) to proactively keep people off their property. A store owner need not wait until a malicious actor enters their unlocked store before informing the actor that they are not welcome and cannot enter. Again, it is a right to exclude, not merely a right to expel. The trespass doesn't occur until after the actor enters, but that does not mean the store owner lacks a right to exclude or is obligated to allow the actor to enter against the store owner's wishes first, or that the store owner's only recourse is to lock the store and block all potential customers from entering, allowing only those who first receive a special key.

---

[140] What Kerr calls "public" is probably better understood as common property, rather than public property. *See* Sec. II(B) *supra*.

[141] *Norms* at 1177, https://www.columbialawreview.org/content/norms-of-computer-trespass/

[142] https://www.columbialawreview.org/content/norms-of-computer-trespass/ at 1151 ("If the door is unlocked, you can enter the store and walk around.")

[143] https://www.columbialawreview.org/content/norms-of-computer-trespass/ at 1153

The more I try to extend the analogy, the more nonsensical it becomes, as it should, because real property is the wrong comparison. It makes more sense to note that just because I bring a backpack of candy into public and allow many people to look through it and grab their favorite sucker, that does not mean I must let people I don't want to access my bag do so before I'm allowed to exercise my property right to exclude. Exclusion is manifestly different from expulsion.

## 4. Internet as a Monolith

On a couple of occasions, Norms refers to the Internet as if it were a single entity, rather than acknowledging that it is composed of millions of independent websites and countless other actors. In this way, using *Norms'* preferred real property analogy, it would be like saying that no stores in a city should have any discretion over who is allowed in. Rather, the focus must be only at the city level.

This framing undermines the autonomy of website owners in the service of a single entity called "the Internet," as he does when noting that his framework creates "public rights to use the Internet."[144] But people do not use "the Internet" any more than they use "the city" they live in. Much like the Internet and websites, cities mostly do not determine how individuals exercise their bundle of rights over their personal property.

While it may be true that many sites are fully open with no restrictions, it does not follow that the Internet requires all websites to be either fully open or fully private. People do not relinquish their right to exclude simply because they choose the best medium to connect with other humans at scale. Similarly, bringing a backpack to a public park does not mean you forfeit the right to exclude anyone from rummaging through it.

It should also not matter what the default setting of the Internet is. *Norms* claims that "when a computer owner decides to host a web server, making files available over the Web, the default is to enable the general public to access those files."[145] And therefore, "Access is inherently authorized." *Norms* goes so far as to declare that "A person who connects a web server to the Internet agrees to let everyone access the computer," which would probably come as a surprise to many. These assertions occur several times, and in each instance, *Norms* does not ask whether a website allows total access, but instead presumes that all "public" websites do, as seen when he begins sentences with phrases such as "because the Web allows anyone to visit…"

Additionally, "the Web" does not allow anything any more than "the roads" allow cars to drive 100 miles an hour. The Web can only enable. It is the websites that determine who can visit and under what circumstances. And though the statements about the default may be generally accurate, that does not mean website owners lose the ability to change it.

---

[144] https://www.columbialawreview.org/content/norms-of-computer-trespass/ at 1161
[145] https://www.columbialawreview.org/content/norms-of-computer-trespass/ at 1162

Yet *Norms* does seem to allow for the slightest wiggle room. The paper notes in passing that "companies that want to keep people from visiting their websites shouldn't connect a web server to the Internet and configure it so that it responds to every request." The text assumes that if the content is public, then the configuration must necessarily allow everyone to access it. But this is precisely what I'm arguing against. Websites can both be public *and* configured to block certain visitors, such as GenAI bots, and they can do so with far more sophisticated means than just IP blocking.[146]

## D. Other Scholarship

*Norms*' focus on the CFAA is not an anomaly. Neither is how Kerr overlooked trespass to chattels despite his paper's title. Ben Sobel's *A New Common Law of Web Scraping* also too quickly skips a proper analysis of trespass to chattels. For instance, Sobel claims that "Under *Hamidi*, a trespass to personal property plaintiff *must* 'demonstrate some measurable loss from the use of its computer system' that is 'substantial,' rather than 'momentary or theoretical.'"[147] (emphasis added)

But that view is too narrow. The court focused on harm to the computer system because it presumed it was the only property at stake. The court did not consider harm to the website's value. In other words, just because an argument was not discussed does not mean that only the arguments that were discussed matter.

Sobel goes on to claim that "As a threshold matter, efforts to mitigate scraping arguably do not 'effectively control[] access' if the webpages remain publicly accessible."[148] But this framing incorrectly implies that web content exists in a binary state: it's either publicly accessible to everyone, or it's not. It also incorrectly implies there is no way to effectively control access such that the content remains public for virtually everyone while excluding certain entities.

Catherine M. Sharkey's paper on self-help and trespass torts is enlightening in many ways, but it, too, suffers from focusing only on facts that have been presented, rather than on how the tort could apply under different facts. For example, when detailing common law trespass to chattels claims, she notes only three possible harms: impairment of a computer server, threat of potential

---

[146] Kerr argues that when someone's IP address is blocked, they are not acting without authorization by changing their IP address to get around the block. His reasoning is that IP addresses change for many innocent reasons. That's fine as far as it goes, but what Kerr doesn't engage with is a scenario when someone intentionally rotates through IP addresses, spoofing locations or pretending to be a residential origin in order to get around the IP blocking. Such intentional actions in order to circumvent unauthorized access when they know or should know their IP address is blocked, is trespass. It can't be the case that if you tell me to stay out of your bag or off your land, I am not legally liable if I return in a disguise that fools you.

[147] *A New Common Law of Web Scraping* at 174, quoting Intel Corp. v. Hamidi, 71 P.3d 296, 306–07 (Cal. 2003).

[148] A New Common Law of Web Scraping at 177, quoting Lexmark Int'l, Inc. v. Static Control Components, Inc., 387 F.3d 522, 547 (6th Cir. 2004)

future harm, and consequential economic damages, such as harm to business reputation and goodwill.[149] None of these encapsulate the harm of a website's value decreasing because the scraping causes significantly fewer people to visit it.

Still other criticisms, such as those by Michael A. Carrier and Greg Lastowka against cyberproperty, seem to focus on uncommon claims.[150] For example, they contend that because one's property online, like a website, contributes so little to the World Wide Web, society must not grant the website *any* property rights.[151] Their stance seems to attack a straw man, claiming that people who want property rights over their online content actually want to control the entire World Wide Web.[152] Perhaps someone has tried to argue that they have the right to control the value of the entire network because they own a website, but that is certainly not a claim made in this paper.

Carrier and Lastowka's argument rests on the idea that "Given the internet's vast network effects, the value of the system far exceeds the value of any individual investment in a single server or website."[153] But this logic would seem to apply to virtually *all* property. The value of a house, a watch, and a patent all derive from the network effects of living in a society. The value of society exceeds the value of any individual human creation within it, yet most people would agree that this does not mean property should not exist.

Other examinations of property on the Internet are even further afield, stating as a premise of the argument that there is an "absence of resource rivalry."[154] But as noted above, this is a misunderstanding of how websites work. The *content* may be nonrivalrous, but *websites* are not. Recall that the trespass is to the website infrastructure, not to the non-rivalrous content it contains.

---

[149] Catherine M. Sharkey, *Trespass Torts and Self-Help for an Electronic Age* at 14-17

[150] I do cut the authors some slack for bold assertions, as their paper was written in 2007. For example, they praise Google's business model (which has since been declared a monopoly for both search and advertising). They similarly missed the mark when they asserted that "'walled garden' models, in which proprietary zones have been segregated from the greater internet, have failed to lead to creativity and innovation. If anything, it is the most heavily "propertized" regimes that have not been capable of long-term survival in the networked ecosystem." Today many of the largest companies are walled gardens: Facebook, Amazon, Tik Tok, etc. Finally, they contend that "Cyberproperty proponents' confidence that the [cyberproperty] doctrine is necessary to prevent a tragedy of the commons is unfounded. There is no evidence that cyberproperty's absence would create a tragedy of the commons," but this paper argues that exactly such a harm is now occurring.

[151] *Against Cyberproperty* at 1499. ("Robert Nozick provided the most famous illustration of this principle. He asked: 'If I own a can of tomato juice and spill it in the sea so that its molecules.., mingle evenly throughout the sea, do I thereby come to own the sea ...?' Obviously not.") They go on to support their argument by noting that online property gains in value when more people use the network and compare it to telephones, which makes one wonder if this mean people should not have a property right in their cell phones either.

[152] *Against Cyberproperty* at 1500 ("even if a chattel owner were to have a Lockean claim over a networked component as a stand-alone object, she cannot claim the value of the entire networked system. Network effects, not individual owners, are primarily responsible for the system's value.)

[153] *Against Cyberproperty* at 1500.

[154] Shyamkrishna Balganesh, *Common Law Property Metaphors on the Internet: The Real Problem with the Doctrine of Cybertrespass*. *See also Against Cyberproperty* at 1503.

## E. Where Critics of Trespass to Chattels Go Wrong

Several standard arguments fall flat under closer scrutiny, and it's time for the courts to reevaluate whether and how trespass to chattels should apply to a website owner's ability to exclude scrapers.

The solution isn't to prohibit all scraping, but to allow each site to decide for itself what can be scraped. If anything, sites have no more of an obligation to share everything on an all-or-nothing basis than any individual in the physical world. Eliminating the right to exclude and retaining only the right to sue for post hoc content use is equivalent to the government granting everyone the right to copy others' work (though such copying may be deemed unlawful at some later point). That can't be right. Some scholars believe it is important to declare that "there is no such thing as absolute property" in the service of arguments that there should be *no* cyberproperty *at all*. But stating that there are no absolutes in property law is merely a truism, and a website owner's right to exclude does not disturb such claims.[155]

The criminalization approach under the CFAA is unnecessary in most cases; intentional torts will suffice. My proposal for exclusion is also beneficial for scraping because, unlike casual users who may be unaware of how websites load and are not prolific scrapers, sophisticated scrapers (the GenAI companies and the companies they pay to scrape data on their behalf) understand how websites function. For an unsophisticated scraper, the blocking will simply prevent scraping, and they won't understand why. For a sophisticated scraper that intentionally circumvents technical safeguards, we must hold it to a higher standard. Importantly, intentional actions invite punitive damages.

If you have a right to exclude, when should that right apply? It must apply before the content can be scraped; otherwise, it's at best difficult to enforce. It's not enough to have the right to *want* to exclude someone. It's a right to *actual exclusion*. The right to exclude must be meaningful, backed by the force of law to discourage attempts to take data. The answer can't be that, as long as someone takes your data against your will, even gently or politely, it is somehow no longer unlawful as a matter of trespass.

While older trespass cases focused on harms to the function of servers,[156] there is no principled reason the harms must focus on the hardware. The effect of scraping on the website's ability to continue existing in a meaningful and useful form matters every bit as much as being disrupted by spam.

---

[155] *Against Cyberproperty* at 1498.

[156] The harm to servers was often considered a form of possession. But dispossession is often an unnecessarily high bar. If a person is wearing a backpack, can another person go behind them and unzip it and look through it without authorization and without trespassing because you didn't dispossess them of the backpack? Of course not.

In almost every case, the value of a website is derived entirely from its content. Furthermore, there is robust evidence that GenAI scraping can harm websites.[157] There is no equally robust counterargument that GenAI generally enhances the value of scraped websites or that it leaves them unaffected.[158]

While there is no dispute that servers are property, the website is not the server. They are severable. The property at issue is the website, not the server on which it resides.

Website owners who lose traffic and, as a result, revenue when their business model relies on ads, subscriptions, sales, or brand or reputation building certainly see their website's value decline. This single factor explains the widespread prevalence of GenAI scraping. Ignoring this would be like saying thieves who frequently steal from a store, thereby giving it a reputation for being a dangerous location, are not violating the store's value; they are merely potentially affecting the value of the particular items they steal.

The catch is that for this form of exclusion to apply to a site, the site must make sufficient efforts to exclude bots.

### 1. The Narrow View

For over twenty years, the prevailing notion of trespass on websites has been that no claim can succeed unless there is harm to the website's technical functioning. For example, the harm might be damage to the server or an overload of the website. However, the harm standard has been too narrowly construed, reflecting a time when relatively few sites were scraped by anything other than web indexing bots.

Recall that the Restatement states, "A trespass to a chattel may be committed by intentionally…using or intermeddling with a chattel in the possession of another." Comment e notes that the trespasser's actions must be "harmful to the possessor's materially valuable interest in the physical condition, quality, or value of the chattel" and that "Sufficient legal protection of the possessor's interest in the mere inviolability of his chattel is afforded by his privilege to use reasonable force to protect his possession against even harmless interference."

Comment h reaffirmed that physical harm is not necessary, stating that "There may, however, be situations in which the value to the owner of a particular type of chattel may be impaired by dealing with it in a manner that does not affect its physical condition.... In such a case, the intermeddling is actionable even though the physical condition of the chattel is not impaired."

In other words, a claim for trespass to chattels should succeed where a website can show intentional intermeddling with the website that caused harm to its value.

---

[157] *See* Sec. VI *supra*.

[158] While the absence of evidence may not be evidence of absence, it is remarkable Google, Meta, OpenAI, and others are unable or unwilling to provide evidence supporting a counterargument.

## Intentionality

First, we can deal with the intentionality standard. The intentionality at issue here is *not* whether the scraping was intentional, as it almost certainly was. Moreover, many websites *want* to be scraped. The question of intentionality is whether the scraper intentionally circumvented technical measures meant to prevent scraping. Given the relatively open nature of the Internet and the inability of websites to convey who is permitted to enter a website via text or imagery *before* an entity enters, it makes sense that, unlike with most personal property, trespassing on a website would require some other efforts to exclude.[159]

Trespass requires two elements: active exclusionary measures by the website and intentional circumvention by the scraper.[160] As with virtually all instances of intentionality, the intentionality of the scraper can be assumed based on certain actions. Merely scraping a site may not be sufficient to show sufficient intentionality. If a website wants to be scraped and fails to implement measures, the scraper should not be punished arbitrarily because the site owner decides to change their mind internally or subjectively. In other words, the measures must be objective. The intentionality standard also means that using robots.txt would likely not suffice to trigger a trespass claim, as it is a passive form of enforcement, and bot deployers could unintentionally skip the file during scraping.

In contrast, if a scraper goes out of its way to circumvent technical measures that cannot otherwise be circumvented by accident, we can presume it did so intentionally, and this could trigger a viable trespass claim.

What follows is a non-exclusive list of actions that could make the tort enforceable: rotating user-agent names to make it challenging to identify the scraper (this is especially true if the site has already explicitly blocked one user-agent from the same company; for example, if a website blocked Anthropic-AI, Anthropic cannot presume that spinning up a Claudebot means it's acceptable to scrape the same site[161]), using a virtual private server to rotate IP addresses so the site can't recognize the true source of the scraper, or spoofing as residential IP addresses.

## Intermeddling

There appears to be little dispute that unwanted scraping bots intermeddle with a website. To intermeddle merely means to interfere. More specifically, it's "To interfere wrongly with property or the conduct of business affairs officiously or without right or title."[162] For the intermeddling to be wrong, it generally needs only to be without consent.[163] This is the standard for personal property writ large.

---

[159] To stick with backpack examples: one does not need to zip their bag in order to reserve a trespass claim for a stranger who decides to rummage through it without authorization.

[160] While dozens of pages have been written about why self-help like technological measures fit into legal theory (*see, e.g.,* Catherine M. Sharkey's *Trespass Torts and Self-Help for an Electronic Age*), the reason I propose it as a requirement for the tort is more practical: without such measures scrapers cannot reasonably know whether they are being excluded.

[161] https://www.404media.co/websites-are-blocking-the-wrong-ai-scrapers-because-ai-companies-keep-making-new-ones/

[162] https://thelawdictionary.org/intermeddle/

Websites are personal property; therefore, the consent standard must apply equally to websites. While many will likely quibble with the idea that scraping is "wrong," the legal arguments generally accept that the bots scrape against the will of the website owners, and the only real question is whether the harm is sufficient to make the intermeddling legally cognizable.

## **Harm**

The harm standard is interesting because, for over 20 years, it has required some form of physical harm caused by bots, making it difficult or impossible for websites to function properly. Again, Sec. 218 says, "One who commits a trespass to a chattel is subject to liability to the possessor of the chattel if, but only if, (a) he dispossesses the other of the chattel, or (b) the chattel is impaired as to its condition, quality, or value, or (c) the possessor is deprived of the use of the chattel for a substantial time, or (d) bodily harm is caused to the possessor, or harm is caused to some person or thing in which the possessor has a legally protected interest."

In other words, courts and scholars have largely or completely overlooked Section 218(b). In particular, they have not considered the harm to website value caused by GenAI scraping.

Section VI demonstrated that GenAI scraping can cause significant harm to websites by extracting and exploiting their content.[164] The bots curtail many sites' ability to gain subscribers, sell merchandise, or even build a brand. The harm is both directly and indirectly attributable to GenAI bots, as evidenced by the rapid decline in website traffic experienced by several sites since the debut of ChatGPT in November 2022. In short, harm to the economic value of a website, including loss of traffic, advertising revenue, or brand reputation, can be just as significant as physical impairment.

Importantly, the harm need not be profound for a claim to succeed. There is no million-dollar threshold for trespass to chattels, and there certainly should not be a special, extra-high bar for websites compared with other forms of personal property. For some, scraping may pose an existential risk even if the absolute amount is relatively small. The harm, in other words, doesn't need to be huge. It needs only to be sufficient to constitute injury.

It is worth noting further that whether harm is a requirement for trespass to chattels is a matter of some debate. As Profs. Brady and Stern wrote, one court stated that "it was not a trespass when a defendant 'rummage[d] through plaintiff's books, papers and records, making voluminous notes' because the plaintiff's property was not 'injured in any way.'"[165] Here is how Brady and Stern conclude that section:

> Despite these few cases on the subject and only minor changes to the Restatement, the famous torts treatise Prosser and Keeton on the Law of Torts pronounced it settled by 1984 that authority, though "scanty," nonetheless supported a conclusion that "the dignitary interest in the inviolability of chattels, unlike that as to land, is not sufficiently important to require any greater defense" than the availability of self-help.[166]

[163] Brady and Stern note the limits of norms and implied consent. ("There are other situations where—even if there was some degree of custom or implied permission at the outset—it vanishes once the owner has made clear their objection and given the person an opportunity to desist.) Other cases supporting this notion include *Bruch v. Carter*, 32 N.J.L. 554, 555 (1867); *Graves v. Severens*, 40 Vt. 636 (1868); and No. 48119, 2023 WL 2563783, at *1 (Idaho Mar. 20, 2023).

[164] This includes, for example, loss of revenue and loss of human visitors.

But as Sharkey points out, "Frederick Pollock's position was 'that in strict theory it must be a trespass to lay hands on another's chattel whether damage follow or not.' Prosser and Keeton's torts treatise, moreover, cites several trespass to chattels cases from the nineteenth and early twentieth centuries in which actual harm to the chattel is not required."[167]

Brady and Stern provide another example, where "in at least one case on trespass to chattels before the Intel decision—*Bankston v. Dumont*—the court found sufficient proof of trespass to chattels in part because a container was violated: a purse."[168] They go on to write that "The scattered cases both before and after the Restatements hinted at the possibility that even absent actual damage, trespasses to chattels might be actionable if the intermeddler intentionally or wantonly contacted the thing, knowing they had no permission to do so."[169]

Although in a different context, some Fourth Amendment cases have also suggested that containers, such as backpacks and websites, should be given greater protection than items without such properties. For example, *United States v. Chadwick* held that individuals have a greater expectation of privacy for containers in a car than for the vehicle itself, meaning a warrant is generally required to search such containers.[170]

In sum, while many website owners will likely be able to demonstrate harm, it is far from clear that they must do so to prevail on a trespass to chattels claim.

### 2. Designed for Dissemination

Public dissemination does not negate exclusion rights. While there are countless examples, we need look no further than news sites.

---

[165] Maureen E. Brady and James Y. Stern, *Analog Analogies: Intel v. Hamidi and the Future of Trespass to Chattels* at 215.

[166] Maureen E. Brady and James Y. Stern, *Analog Analogies: Intel v. Hamidi and the Future of Trespass to Chattels* at 216.

[167] Catherine M. Sharkey, *Trespass Torts and Self-Help for an Electronic Age* at 127

[168] Maureen E. Brady and James Y. Stern, Analog Analogies: Intel v. Hamidi and the Future of Trespass to Chattels at 218 citing Bankston v. Dumont, 38 So. 2d 721, 722 (Miss. 1949).

[169] Maureen E. Brady and James Y. Stern, Analog Analogies: Intel v. Hamidi and the Future of Trespass to Chattels at 221.

[170] 433 U.S. 1 (1977)

It would be odd to think that the *Wall Street Journal* or *New York Times*, which spend millions of dollars on journalists, equipment, real estate, travel, and more, should have no right to exclude any scrapers, including AI scrapers that repackage news site information and give it away for free, because they make some content public for SEO and to attract potential subscribers. In other words, they do so to survive as a business. There is no dispute that newspapers create content to inform public discourse, hoping their articles will be widely read. There is also no dispute that making a copy of a physical newspaper without authorization (especially when the purpose of the copying is to benefit the copier financially), even if the newspaper is publicly viewable on a newsstand, would be unlawful. Yet now we are in a space where arguments about what information must be accessible hinge on whether the information can ever be accessed without first logging in to an account.

Opponents of exclusion seem to misunderstand the economics of many sources of "publicly available" information. Sometimes, a site lets visitors access some content for free so they can get a taste and decide whether to subscribe. For example, a site might allow three free articles a month before requiring payment. Other sites make information free in the hope of generating advertising or affiliate marketing revenue. Still others do it to build a reputation and become associated with the place where certain ideas and products originate. To make this point more explicit, in no case does a serious for-profit news business share information freely without any hope of generating revenue; yet that must be the worldview of those who oppose a right to exclude, given how websites share some information publicly.

### 3. Stifling Expression and the Exchange of Ideas

Finally, opponents of the right to exclude who argue that exclusion would stifle expression have it precisely backward. It's not that blocking scrapers will stifle speech; it's that prohibiting exclusion will. When people create content, they do so to reach others. People do not write stories, essays, or songs for the enjoyment of bots. And while some may argue that bots are merely intermediaries that connect people interested in content with that content, this overlooks the fact that when GenAI paraphrases content, it often omits attribution to the source material, and even when it includes attribution, far fewer people click on links than those who encounter the same information via traditional search.

When websites are left largely defenseless due to the weakening of trespass-to-chattels claims, they may resort to hiding content behind paywalls and login screens, seeking refuge in the CFAA. Alternatively, they might block search-indexing bots, such as Googlebot. This is because Google uses its bot not only to generate search results but also to create AI Overviews and snippets. This makes information harder to find and has the side effect of chilling speech. Instead of encouraging the creation and sharing of ideas, allowing bots to scrape public websites leads to self-censorship.

Moreover, the content generated by GenAI through internet scraping does not promote science or the arts in a manner that approaches the profound insights and creations of humans. General-purpose GenAI has not made a significant scientific discovery or insight, nor has it created a new genre of art. Instead, GenAI often leads to what is now commonly referred to as AI slop, or low-quality, usually formulaic and repetitive, content generated by artificial intelligence, characterized by a lack of effort, originality, and factual accuracy, and is often produced in large quantities. If you have used the Internet for five minutes in the last three years, you have likely encountered AI slop.[171] This should make the tradeoff of allowing scraping in exchange for GenAI even harder to swallow.

In short, there are several reasons why revitalizing trespass to chattels for scraping makes sense legally and practically. Allowing the claim encourages speech, promotes science and the arts, and supports business models that have been proven to work (unlike the unprofitable business models of GenAI-only companies), thereby benefiting the economy. In contrast, GenAI scraping tends to be parasitic, harvesting others' personal data and work without the creators' consent and without any clear, proven, or overwhelming benefit to society.

Having addressed the main criticisms of revitalizing trespass to chattels, the next logical question is: what concrete steps can website owners take to protect their interests in the current legal and technological landscape?

# IX. What's a Website to Do?

Websites are personal property; thus, it makes sense to treat them as such, including granting the owner all the rights of a property owner. This includes the right to exclude and the ability to sue for trespass when scrapers circumvent the website owner's wishes.

If there is a right to exclude, what is required for that right to be legally cognizable? Because the internet, unlike the physical world, often facilitates interactions between total strangers, we usually cannot presume that site visitors understand who the website owner does and does not want on the site. Mere consideration of who should not be allowed cannot be sufficient to raise a trespass claim when the person or entity being blocked cannot reasonably know they were meant to be excluded. Instead, the law should require claims to show that a trespasser intentionally circumvented a sufficiently knowable barrier.

## A. Sufficient Efforts of Technical Exclusion

For the right to exclude to apply, the website must take active steps to prevent its site from being accessed by scrapers.[172] Note that the focus is on access, not on preventing scraping *after* the scrapers gain access to the site and have therefore very likely already copied its contents.

To bolster their defenses, website owners are increasingly using advanced bot protection tools beyond robots.txt. These include mandatory sign-up and login requirements to restrict content access to authenticated users, as well as security services such as Cloudflare Firewall and AWS

[171] This is, of course, a little hyperbolic. But less so with each passing day.
[172] If the content is scraped before the website implements these technical measures, a trespass to chattels claim is unlikely to succeed.

WAF for real-time bot detection and blocking, IP blocking, HTTP header analysis, and JavaScript-based fingerprinting.

Tollbit's Q2 2025 report suggests that the efforts at blocking AI bots is not futile:

> Over the last year, 302 [temporary redirect], 402 [payment required], 403 [access forbidden/blocked], and 401 [not authenticated] response statuses have increased by almost 4x on Tollbit partner websites, likely due to publishers deploying security measures (bot detection/WAF) that prevent non-human visitors such as AI bots from accessing content.[173]

These technical measures are critical because they create barriers that, if bypassed, can significantly strengthen trespass claims even where they may not be sufficient for criminal statutes. Many are also cheap or free, and easy to implement.[174] More importantly for this paper, they also provide concrete evidence of a website owner's intent to exclude and can bolster arguments for unauthorized interference in trespass-to-chattels claims. As Professor Sharkey notes, self-help can serve "as a 'sincerity index'" and "as a way in which someone can 'mark' his property as private–or exclude it from the public commons."[175]

Finally, it may be that courts view such technological barriers as a necessary element. The court in *CompuServe* suggested self-help was a prerequisite to filing a trespass claim, stating that "[T]he implementation of technological means of self-help, to the extent that reasonable measures are effective, is particularly appropriate . . . and should be exhausted before legal action is proper."[176] The same court noted that "[W]here defendants deliberately evaded plaintiff's affirmative efforts to protect its computer equipment from such use, plaintiff has a viable claim for trespass to personal property . . . ."[177]

This paper does not rely on a particular type of barrier or prescribe what satisfies the trespass threshold. The test is whether any bypassing of barriers to access content on a website was intentional, and due consideration must be given to whether the barriers are robust enough that normal attempts to access the website would be blocked, so that the only way around them is to deploy sophisticated circumvention methods. Arguments about sufficiency need not require a deep technical exploration by the trier of fact. The common sense and lived experience of an ordinary judge or jury will usually suffice.

The "casual surfer" test can serve as a simple guide: If a bot can access the content under circumstances similar to those of a casual web surfer (even for a fraction of a second), even if it requires some effort that is reasonable (such as clearing her cache, using incognito mode, or using a different browser), then the content is probably not sufficiently protected against the bot to support a viable trespass to chattels claim. In contrast, if the bot cannot access the content in the manner of a casual surfer despite taking those common extra steps, then the barriers are likely sufficient.

---

[173] TOLLBIT, *AI Scraping Is On The Rise: TollBit State of the Bots - Q2 2025*, (Sept. 8, 2025), https://tollbit.com/bots/25q2/.

[174] Setting up Cloudflare's bot blocking from scratch takes no more than 15 minutes and is free, for example.

[175] Catering M. Sharkey, *Trespass Torts and Self-Help for an Electronic Age* at 8.

[176] 962 F. Supp. 1015 at 1023

[177] 962 F. Supp. 1015

In short, the test is whether the scraper circumvented a sufficient technical barrier designed to block it (such as a WAF or IP block) in a way an ordinary person would not. It requires (a) a demonstrable, good-faith effort to exclude someone or something (such as bots) using reasonable technical means and (b) proof of the deliberate bypass of that exclusionary effort. It does not require proving that the site is inaccessible to all by default.

Finally, it is important to recognize the ongoing technological arms race between bot developers and site administrators. No barrier is impenetrable. The legal standard should require reasonable, good-faith efforts, not absolute security. This approach aligns with established tort principles, which require only reasonable care under the circumstances, and ensures that property rights remain meaningful in a rapidly evolving digital environment.

## B. Cloudflare as an Example of a Sufficient Barrier

Some companies that help websites identify and block AI bots before they can access content include Fastly, Datadome, and Cloudflare. This section will use Cloudflare's product as a representative example because it has emerged as a key enabler of these defensive strategies. In September 2024, Cloudflare introduced a one-click option to block AI crawlers, an option chosen by over one million customers.[178] By July 2025, Cloudflare took a significant step further by blocking unauthorized AI crawlers by default for all new domains upon sign-up, requiring explicit permission from website owners for AI companies to scrape content.[179] This initiative includes the launch of "Pay Per Crawl," a monetization tool allowing publishers to charge AI firms for data access. Early adopters of these measures include major publishers like Gannett, Time, and Stack Overflow.

Cloudflare's technical capabilities demonstrate that bot exclusion is both feasible and effective. The way it detects and blocks bots reveals the extent to which a site can exclude them. Namely, that it is possible to do so, and there is no reason to assume bots can or should gain access to sites just because ordering people can. Cloudflare intervenes to block AI scraping bots at the earliest stages of the request, before the webpage begins loading on the client's (bot's) side. Specifically, Cloudflare's bot management operates at the network edge, intercepting traffic before it reaches the website's origin server and before the browser initiates HTML parsing and rendering.

---

[178] Press Release, Cloudflare, Cloudflare Just Changed How AI Crawlers Scrape the Internet at Large (July 1, 2025), https://www.cloudflare.com/press-releases/2025/cloudflare-just-changed-how-ai-crawlers-scrape-the-internet-at-large/.

[179] Press Release, Cloudflare, Cloudflare Just Changed How AI Crawlers Scrape the Internet at Large (July 1, 2025), https://www.cloudflare.com/press-releases/2025/cloudflare-just-changed-how-ai-crawlers-scrape-the-internet-at-large/.

Another way to think of it is that Cloudflare sits "in front" of websites (a space whose existence many courts seem unaware of). All traffic to the websites using Cloudflare's services first passes through its global network. They then employ a multi-layered approach to bot detection, including heuristics,[180] machine learning,[181] IP reputation,[182] JavaScript detections,[183] and user-agent analysis.[184] Note that these sophisticated methods are beyond the capabilities of most homemade websites. The local Mom-and-Pop cafe's website likely does not use machine learning, for example, unless it relies on a service like Cloudflare, Fastly, or others.

Based on the supply chain steps described in Section IV, Cloudflare's intervention occurs primarily during steps 3 and 4. As previously established, bot scraping occurs no earlier than step 5. Consequently, Cloudflare effectively excludes those bots.

To recap, step 3 involves establishing a connection between the browser and the server (the website location) through the TCP/TLS handshake. Even if the handshake completes, Cloudflare can immediately assess the nature of the incoming connection and decide whether to proceed.

Step 4 involves the browser sending an HTTP request to the server to retrieve the website's content. This is perhaps the most critical stage for Cloudflare's intervention. Cloudflare's systems analyze the HTTP request itself and associated metadata (IP address, user-agent, JA3/JA4 fingerprints,[185] behavioral patterns, etc.) to determine if it's a legitimate request or a bot.

Cloudflare can accurately exclude bots in part because it has access to a tremendous amount of data by virtue of approximately a fifth of all internet traffic flowing through Cloudflare's architecture.[186] This vast dataset enables the continuous refinement of detection algorithms and the rapid identification of emerging bot patterns.

---

[180] E.g., analyzing known malicious patterns and signatures.

[181] Machine learning models analyze traffic data to identify anomalies and distinguish between human and bot behavior (e.g., mouse movements, click patterns, and interaction times). Jin-Hee Lee et. al, *Control Content Use for AI Training with Cloudflare's Managed robots.txt and Blocking for Monetized Content*, Cloudflare Blog (July 1, 2025), https://blog.cloudflare.com/control-content-use-for-ai-training/.; "What is Bot Traffic?," *Cloudflare Learning*, https://www.cloudflare.com/learning/bots/what-is-bot-traffic/ (Anomalies used to identify bots include abnormally high pageviews, abnormally high rate of visiting a site without clicking on anything, surprisingly high or low session duration, junk conversions, and spikes in traffic from unexpected locations.) The sheer volume of internet traffic that Cloudflare facilitates–an average of 57 million requests per second–allows it to refine and update its models to identify bad actors based on patterns or trends.

[182] E.g., blocking known malicious IP addresses or ranges.

[183] Which involves injecting lightweight JavaScript challenges that legitimate browsers can easily execute but many headless browsers or simple scrapers cannot.

[184] E.g., identifying suspicious or blacklisted user-agent strings.

[185] JA3 and JA4 fingerprints are a method for identifying the client-side software used in TLS (Transport Layer Security) communication, based on the specific cryptographic parameters they send during the TLS handshake. These fingerprints allow network defenders to classify and detect malicious or unusual network traffic by recognizing the "fingerprint" of the application making the connection, regardless of its IP address, domain, or what it is trying to access.

[186] This also means that its efforts to block AI bots only work if the websites use Cloudflare's products.

## C. Log-in Pop-ups as an Example of an Inadequate Barrier

Sometimes, sites load, and then, within a second, a pop-up appears over the content, obscuring the background and requiring a login to access it.

This pop-up can occur at different steps in the webpage loading process, depending on several factors and owner preferences (e.g., search engine optimization, ease of implementation, aesthetics). However, the most common points for this approach are after the initial content (or a placeholder) has loaded. This would be after step 6.

Step 6 is when HTML parsing and DOM construction occur. Therefore, the pop-up does not appear until the browser receives instructions to render the site. The pop-up is delayed until after step 6 because many websites want to display something to the user, even if it's just a blurred version of the content or a "teaser" of what's behind the login wall.

The HTML and basic CSS are usually loaded and parsed first. Then, the JavaScript code (downloaded in step 7) runs to check the user's login status. If the user isn't logged in, the JavaScript dynamically displays the login pop-up. This allows the website to render quickly, and the pop-up appears after the user has a sense of what the page is about.

As an alternative, the pop-up can appear during step 11, when JavaScript executes (if the content is dynamically loaded or hidden). Some websites load very little content initially and rely heavily on JavaScript to fetch and display the page content. In such cases, the JavaScript that determines the login status and displays the pop-up would run as part of the overall JavaScript execution. The pop-up might appear before all dynamic content is loaded, preventing the user from seeing anything meaningful until they log in.

Why don’t the sites display the pop-up sooner? The answer lies in the technical constraints of web page loading. Steps 1-4 are too early. The browser hasn't received the HTML yet, so it can't display a pop-up. The server might redirect to a login page instead of the original content, but that's a full-page redirect, not a pop-up on the current page. Unfortunately, waiting until later to initiate the pop-up allows bots to scrape the content before the pop-up blocks them.

If the site has not implemented measures to block bots before the pop-up (i.e., before step 5), bots can still likely scrape the content *even if the site uses a pop-up*.[187] In other words, login screens and paywalls may not be effective at excluding scrapers. This may help explain why courts have historically presumed that bots have access to all content and tended to start their analysis after step 5. Websites have historically employed mostly ineffective methods of blocking

[187] This is not to say that websites that offer a few free articles necessarily surrender their ability to bring trespass claims. But it does mean they will need to use more robust technical measures than just a pop-up login screen to strengthen their claims.

bots, leading courts to incorrectly assume that bots cannot be blocked from accessing "public" content.[188]

## D. Perplexity's Efforts as an Example of Trespassing

Finally, it's helpful to see a clear example of what would be trespass to chattels. For this illustration, I will use Perplexity's actions during content scraping.

Cloudflare publicly named and (attempted to) shame Perplexity for its aggressive scraping methods in August 2025. In a blog post, Cloudflare detailed how Perplexity obscures its crawling identity when presented with a network block, thereby circumventing the website owner's preferences and scraping the underlying content. In contrast to the generally accepted best practices detailed in Section V, Perplexity modifies its user agent, changes its autonomous system number, and ignores or fails to check robots.txt.[189]

Perplexity also uses headless browsers, which are "intended to impersonate Google Chrome on macOS when their declared crawler was blocked."[190] The "undeclared crawler utilized multiple IPs not listed in Perplexity's official IP range, and would rotate through these IPs in response to the restrictive robots.txt policy and block from Cloudflare."[191]

OpenAI stands in stark contrast to Perplexity's activities. According to Cloudflare, OpenAI identifies and describes its bots on its website, respects robots.txt, and does not appear to evade blocking. If accurate, this shows that a company can be a leader in GenAI while still respecting website owners' personal property rights.

While these strategies offer website owners new avenues for protection, it is important to recognize the limitations and challenges inherent in enforcing exclusion rights online.

# Part 3: What Follows

Suppose that trespass to chattels is reinstated as an enforceable tort. What then? Part 3 will examine the limitations it entails, possible exceptions grounded in public policy considerations, and anticipated counterarguments opposing the reinstatement of trespass to chattels as a viable means of controlling access to one's website.

---

[188] This legal presumption is also likely how Common Crawl was able to scrape the *New York Times* for several years, despite the *New York Times* having a paywall pop-up.

[189] Gabriel Corral et al., *Perplexity is using stealth, undeclared crawlers to evade website no-crawl directives*, Cloudflare Blog (Aug. 4, 2025), https://blog.cloudflare.com/perplexity-is-using-stealth-undeclared-crawlers-to-evade-website-no-crawl-directives/.

[190] Gabriel Corral et al., *Perplexity is using stealth, undeclared crawlers to evade website no-crawl directives*, Cloudflare Blog (Aug. 4, 2025), https://blog.cloudflare.com/perplexity-is-using-stealth-undeclared-crawlers-to-evade-website-no-crawl-directives/.

[191] Gabriel Corral et al., *Perplexity is using stealth, undeclared crawlers to evade website no-crawl directives*, Cloudflare Blog (Aug. 4, 2025), https://blog.cloudflare.com/perplexity-is-using-stealth-undeclared-crawlers-to-evade-website-no-crawl-directives/.

# X. Limitations

Some significant limitations remain. For one, the right to exclude would still require websites to take affirmative steps to control access to their content. Websites cannot simply make their content publicly accessible without restriction and claim a right to exclude; at most, they may have limited remedies after the fact.[192] Moreover, the steps the website takes must be meaningful. Simply typing "do not enter" on a webpage is not enough. The steps must be sophisticated enough that the only way for an undesirable entity to access the content is by intentionally subverting the technical measures in place.

Another limitation is that individuals who post on a website they do not own are at the mercy of the website owner to protect their content. For example, those who post on Reddit likely do not have the right to exclude GenAI scrapers. Instead, Reddit must take steps to exclude GenAI scrapers at the website level. It may therefore be in the website's best interest to implement and promote such technical measures to reassure content creators that their content will be protected from scrapers. Ideally, this would create a strong market effect, encouraging other websites to compete to protect their creators' content.

A final limitation is that allowing websites to exclude scrapers could encourage the monopolization of information. This was a concern of the court in *hiQ v. LinkedIn*, for example. However, this is a problem for other areas of law, such as antitrust and business competition, rather than property law. Notably, monopolization concerns would likely apply only to a handful of the world's largest websites. A similar concern arises when platforms block scrapers from user content that the user wants to be public, which could spark First Amendment free-speech arguments, but that is a matter for the First Amendment.

Beyond practical limitations, there are also important exceptions to the right to exclude, particularly when public policy considerations require broader access to information.

# XI. Exceptions to the Right to Exclude

Exceptions should be recognized when public policy demands access. For example, following Daniel Solove's recommendations, exceptions might include scraping for socially valuable purposes such as academic research, journalism, public health, or government transparency. These carveouts would prevent the right to exclude from being applied too rigidly and ensure a balance between property rights and broader societal interests. Importantly, "creating GenAI" likely would not suffice for an exception because it is far from clear that GenAI's societal benefits outweigh its harms at this time.

---

[192] E.g., copyright infringement, intrusion upon seclusion, unjust enrichment, breach of contract, etc.

To be sure, GenAI technologies have the potential to drive innovation, improve access to information, and support new forms of creativity. However, these benefits must be weighed against the risks of uncompensated exploitation and harm to content creators (not to mention deepfakes, misinformation, cyberattacks, fraud, and so on). They should not be presumed to outweigh the interests of website owners without rigorous, evidence-based analysis. Policymakers should regularly review the impact of GenAI on innovation, education, and public welfare. Where clear, demonstrable benefits exist, narrowly tailored exceptions to exclusion rights may be warranted. Until then, the precautionary principle counsels in favor of robust protections for website owners.

As Solove puts it:

> [A]rticulations of what constitutes the "public interest" should be specific, compelling, grounded in reality, and directly related to the collection of information. Mere conveniences such as workplace efficiencies or more seamless commercial transactions should not qualify. Mere allegations that scraping will help "keep people safe" or "improve your health" should be insufficient without convincing proof that a demonstration that the scraping is necessary and proportionate to the purpose. Industry will likely attempt to dilute and work around any rule in order to maximize profit, and lawmakers should craft their rules accordingly.
>
> Another factor in the analysis should be whether AI models trained on scraped data were created with better public involvement. Essentially, scraping would be understood as a special privilege to be allowed when certain conditions exist. In order for AI development with people's data to be permitted, individuals or the public should receive something in return. This is a kind of grand bargain, a wide-scale compromise of people's privacy in exchange for something that benefits people, not just a way for companies to make a profit.
>
> …More importantly, those benefits must be proportional to or exceed the benefits to the scraper. Too often, companies will offer some modest or trivial benefit like a mild efficiency for queues or organization as a pretext for information extract that is lucrative only for the scraper. Other times, companies want to scrape so they can offer an important-sounding benefit that in practice is either illusory or so abstract as to be meaningless. "Keeping people safe" is a virtuous goal, but without so many AI surveillance systems don't meaningfully provide safety to society and certainly not to marginalized and vulnerable groups like people of color who feel the brunt of surveillance first and hardest.[193]

In short, there are likely several instances in which circumventing technical measures to scrape the underlying content may be lawful, as the scraping serves the public interest.

[193] https://papers.ssrn.com/sol3/papers.cfm?abstract_id=4884485

However, the onus is on the scraper to make that argument, and the threshold to satisfy such claims must be substantial.[194]

To ensure that exclusion rights do not unduly hinder socially valuable activities, courts and policymakers should adopt a balancing framework for exceptions. Factors to consider include the purpose of the scraping, the availability of less intrusive alternatives, the potential harm to the website owner, and the societal benefits of access. This approach mirrors the fair use doctrine in copyright law, which weighs multiple factors to determine whether unauthorized use is justified. By articulating clear criteria for exceptions, the law can protect both property rights and the public interest.

Even with exceptions in place, some may still raise additional objections to strengthening exclusion rights. The following section anticipates and responds to these arguments.

# XII. Anticipating Arguments

Legal theorists have advanced a handful of arguments for why society should not grant websites the full powers of personal property, including a robust version of the right to exclude. This section will briefly consider each of them and identify their flaws.

## A. Argument 1: If we allow websites to charge for crawling, only the wealthiest GenAI companies will be able to afford the data.

This paper rejects the argument that requiring payment for something means only the wealthy can afford it. That logic, if accepted, would undermine the very notion of property rights. Society does not provide goods or professional services for free simply because some individuals cannot afford them. Likewise, creators should not be compelled to hide their content or allow everyone to access their property, thereby surrendering control of their intellectual property. Similarly, others should not be forced to surrender personal data to GenAI companies without compensation. If cost is truly a barrier and access is truly essential, the government can establish grants to help entities acquire the necessary data.

## B. Argument 2: If a website blocks AI scrapers, it will be hurting itself, because people are increasingly using GenAI as a search engine.

Early evidence suggests that GenAI search generates minimal referral traffic compared with traditional search engines. Thus, exclusion may not meaningfully reduce engagement and may, in fact, protect sites from uncompensated free riding.

[194] To date, studies do not consistently demonstrate that GenAI is producing a large return on investment for businesses, improving education, or is benefiting the environment. It's far from certain it ever will.

## C. Argument 3: Rather than allow a sweeping right of exclusion, we should just allow an option to implement pay-per-crawl.

Pay-per-crawl is an increasingly viable alternative.[195] Some approaches, such as Tollbit and Cloudflare's, allow sites to choose whether to enable bots to scrape their site, charge bots before they can scrape, or block bots. While not a panacea, it can offer creators greater control over how their content is accessed and monetized. By letting website owners set terms and receive compensation, these models may help balance the interests of creators, users, and GenAI companies.

While this may satisfy publishers focused solely on marginal revenue, it reduces creative labor to a low-value commodity. For creators seeking to build a reputation, maintain audience trust, or control brand identity, pay-per-crawl does not address the core harm. The pay-per-crawl argument fails altogether when a website seeks to protect personal information and has no interest in monetizing it.

## D. Argument 4: There are lots of good reasons to allow scraping and to prevent a strong right to exclude, including scraping for scientific research and search indexing.

I agree with this argument. As a reminder, this paper focuses on GenAI bots, not indexing or research bots. Collateral effects may occur; however, empirical data indicate that the overwhelming majority of scraping is commercial rather than research-based. Moreover, traditional indexing by Google, Bing, and others is distinguishable because it is narrowly tailored to promote discoverability rather than to substitute for or displace original works. Notably, websites have always had the option to opt out of being indexed.

## E. Argument 5: Whatever problems that may currently exist, today is the worst these GenAI products will ever be. Issues of referrals and citations may be fixed.

This may be true, but that doesn't necessarily mean much. Just because a product may improve doesn't mean it will significantly improve, improve in ways meaningful to website owners, or that the improvements will arrive quickly enough to mitigate the harms already occurring—such as reduced web traffic, loss of advertising revenue, or diminished incentives to produce high-quality content.

---

[195] *See, e.g.,* Will Allen & Simon Newton, Introducing Pay Per Crawl: Enabling Content Owners to Charge AI Crawlers for Access, Cloudflare Blog (July 1, 2025), https://blog.cloudflare.com/introducing-pay-per-crawl/.

### F. Argument 6: It should not be unlawful to copy publicly available information.

Whether copying information is lawful is a matter for privacy and copyright law. This paper examines whether it's lawful to exclude others from one's property. If the argument is that because some property is publicly viewable, it must be freely accessible to all, that misstates basic property doctrine. A front lawn may be visible to the public, but that does not create a legal entitlement to walk across it. The same reasoning applies to digital property.

### G. Argument 7: Referral traffic to publishers from GenAI searches and summaries may be lower in quantity, but it reflects a stronger user intent compared with casual web browsing.

Assuming this is true, it's still insufficient. This kind of user intent may be effective for funneling and filtering people interested in making a purchase, but it's unclear what "user intent" means for sites that exist for reasons other than selling products. Scholarly, nonprofit, or creative communities derive value from readership, recognition, and cultural participation—not merely high-intent commercial clicks. GenAI summaries divert that value to GenAI developers at the expense of creators.

## XIII. Conclusion

The law does not need to create new protections for websites. It need only apply the existing ones. Websites are personal property. Personal property owners have the right to exclude. When GenAI scrapers circumvent technological barriers to harvest content at scale, they commit trespass to chattels.

The current framework fails on every level. Copyright preemption has forced website owners into a battlefield where the fair use doctrine grants AI companies a presumptive license to scrape. Courts dismiss trespass claims for lack of server damage, ignoring the economic devastation scrapers inflict. Meanwhile, GenAI companies accumulate billions in value while creators subsidize their training data at no cost. This arrangement is not inevitable. It is a choice, and we can unmake it.

Reviving trespass to chattels restores balance. It recognizes that websites are not mere content repositories but integrated digital assets whose value can be destroyed by unauthorized extraction. It acknowledges that circumventing access controls is not clever engineering but unlawful trespass. It provides website owners with meaningful legal recourse, including damages sufficient to deter predatory scraping. Most importantly, it aligns digital property law with physical property law by applying the same exclusionary principles that govern every other form of personal property.

The solution is straightforward: enforce property rights in the digital space with the same rigor we apply to physical property. Treat websites as the chattels they are. Hold scrapers liable when they bypass access controls. Award damages that reflect harm to a website's value, not just server capacity. And trust that property owners, armed with enforceable exclusionary rights, will make better decisions about access than courts issuing categorical pronouncements about the future of AI.

GenAI companies have built empires on other people’s property. Trespass to chattels is not a radical innovation. It is a time-tested remedy for unauthorized interference with personal property. Applying it to websites requires only that courts recognize what website owners have known all along: their digital property deserves the same legal protection as any other form of personal property. The right to exclude is not negotiable. It is foundational. And it applies online.